\documentclass[pre,twocolumn,superscriptaddress,showpacs,longbibliography]{revtex4-1}
\usepackage{graphicx}
\usepackage{epsfig}
\usepackage{amssymb,amsmath,amsfonts,hyperref}
\usepackage{wasysym}
\usepackage{latexsym}
\usepackage{eucal}
\usepackage{verbatim}
\usepackage[caption=false]{subfig}
\usepackage[section]{placeins}
\usepackage[normalem]{ulem}
\usepackage{color}

\newcommand{\be}{\begin{eqnarray}}
  \newcommand{\ee}{\end{eqnarray}}

\nocite{*}
\begin{document}
\title{Quantum cluster algorithm for frustrated Ising models in a transverse field}
\author{Sounak Biswas}
\affiliation{\small{Tata Institute of Fundamental Research, 1 Homi Bhabha Road, Mumbai 400005, India}}
\author{Geet Rakala}
\affiliation{\small{Tata Institute of Fundamental Research, 1 Homi Bhabha Road, Mumbai 400005, India}}
\author{Kedar Damle}
\affiliation{\small{Tata Institute of Fundamental Research, 1 Homi Bhabha Road, Mumbai 400005, India}}
\begin{abstract}
Working within the stochastic series expansion framework, we introduce and characterize a new quantum cluster algorithm for quantum Monte Carlo simulations of transverse field Ising models with frustrated Ising exchange
interactions. As a demonstration of the capabilities of
this new algorithm, we show that a relatively small, ferromagnetic next-nearest neighbour coupling  drives the transverse field Ising antiferromagnet on the triangular
  lattice from an antiferromagnetic three-sublattice ordered state at low temperature to a ferrimagnetic three-sublattice ordered state.

\end{abstract}

\pacs{75.10.Jm}
\vskip2pc

\maketitle
\section{Introduction}
The stochastic series expansion quantum Monte Carlo method~\cite{MelkoRev}, based on
a stochastic sampling of terms in the high temperature expansion of a quantum
many-body system, is among the most useful computational
tools for the study of bosonic many-body systems. When the model being
studied has a locally conserved charge (such as boson number or a component
of the total spin), nonlocal {\em directed loop} updates are possible within this scheme~\cite{Sandvik0,Syljuasen_Sandvik}.
These lead to very efficient sampling of the high-temperature expansion
and allow for a study of fairly large system sizes.

When there is no locally conserved charge, a different
strategy needs to be used. This has been explored previously in the context of
Ising models in a transverse field~\cite{Sandvik}. In this case, a quantum cluster
algorithm has been formulated and tested in some examples~\cite{Sandvik}. Our motivation here is the following: In a broad class of frustrated Ising models in which the application of a transverse field leads to interesting physics, we have found that the clusters grown by this quantum cluster algorithm do not have ``nice'' statistical properties in the sense that they are either too small to be of much use, or too large, leading to updates that are nearly global spin-flips of the entire configuration. This leads to a very significant deterioration of ergodicity,
and affects our ability to obtain reliable and accurate data at large sizes in interesting
parameter regimes.

This motivates our introduction of a new quantum cluster algorithm for Quantum Monte Carlo simulations within the stochastic series expansion framework. In this article, we describe this new algorithm and demonstrate
that it allows for an efficient study of such frustrated Ising antiferromagnets in a transverse field. In order to facilitate a detailed discussion of our method, we focus on a particular two-dimensional model system in the rest of this article, although the underlying idea has much wider applicability, and is not restricted to two dimensional systems. The particular model system we use to illustrate our scheme is the
transverse field Ising antiferromagnet on the triangular lattice with additional further neighbour
interactions.

In this model system, we provide two illustrations of the utility of our new method. First, 
we study the 
antiferromagnetic three-sublattice ordered low-temperature phase~\cite{Isakov_Moessner} of the transverse field Ising antiferromagnet on the triangular lattice, and demonstrate that our algorithm allows us to access key low temperature properties that
are out of the reach of the standard~\cite{Sandvik} quantum cluster approach. Second, we use our method to demonstrate
that a rather weak additional ferromagnetic coupling between next-nearest neighbours drives a phase transition to a ferrimagnetic three-sublattice ordered state at low temperature, and study the performance of our algorithm in this challenging regime of parameter space in which the nature of the ground state depends crucially on the balance between quantum fluctuations, which favour an antiferromagnetic three-sublattice ordering pattern, and
ferromagnetic next-nearest neighbour interactions, which favour a ferrimagnetic three-sublattice ordering pattern.

The rest of this paper is organized as follows: In Section~\ref{Model}, we provide an introduction to the physics of the
transverse field Ising antiferromagnet on the triangular lattice, and summarize our understanding
of the phase diagram of this model system. In Section~\ref{Method}, we review the standard quantum cluster algorithm~\cite{Sandvik} and then describe our new algorithm. In Section~\ref{Results}, we first introduce the physical observables of interest to us, as well
as the measures we use to characterize the performance of our algorithm. We then
use these measures to characterize the performance of our algorithm. Additionally,
we display our results for the low temperature transition between antiferromagnetic and ferrimagnetic three-sublattice order in this model system.
We close with a brief discussion of connections to previous work and possible extensions in Section~\ref{Outlook}.

\section{Model}
\label{Model}
The transverse field Ising antiferromagnet on the triangular lattice, with Hamiltonian
\begin{equation}
  H_{\mathrm{TFIM}} = J_{1}\sum_{\langle \vec{r} \vec{r}'\rangle }\sigma^z_{\vec{r}} \sigma^z_{\vec{r}'} -\Gamma \sum_{\vec{r}}\sigma^x_{\vec{r}} \; ,
  \label{Ising}
\end{equation}
where the sum over $\langle r r' \rangle$ runs over links of the triangular lattice, has been studied earlier as a simple, paradigmatic example of the interplay
between quantum fluctuations and frustrating antiferromagnetic exchange couplings~\cite{Moessner_Sondhi_Chandra,Moessner_Sondhi,Isakov_Moessner}.
It provides a simple example of ``order-by-disorder'' effects in the following sense:
when $\Gamma=0$, the classical Ising antiferromagnet
has a macroscopic degeneracy of minimum exchange-energy configurations that give rise to  a non-zero entropy-density in the $T \rightarrow 0$ limit~\cite{Wannier,RMF}. In this classical
limit, the Ising spins remain disordered all the way down to $T=0$~\cite{Wannier,RMF}, albeit
with a diverging correlation length, providing a simple example
of a classical spin liquid. At $T=0$, the spins have power-law
correlations at the three-sublattice wave vector~\cite{Stephenson}.
The transverse field $\Gamma$ induces quantum mechanical transitions that
flip the individual $\sigma^z_{\vec{r}}$. In spite of inducing these quantum fluctuations, a small transverse field immediately stabilizes long-range three-sublattice order of $\sigma^z$~\cite{Moessner_Sondhi,Isakov_Moessner}.

This long-range order is of the {\em antiferromagnetic} variety (with no accompanying
net magnetization)~\cite{Isakov_Moessner}, and should be
contrasted with the {\em ferrimagnetic} three-sublattice order exhibited at low temperature by the classical Ising antiferromagnet with ferromagnetic next-nearest neighbour couplings~\cite{DP_Landau}. The distinction between the two kinds of three-sublattice order is best summarized
by the following caricature: When the ordering is of the antiferromagnetic variety, the system spontaneously chooses
one sublattice (out of the three sublattices corresponding to the natural tripartite decomposition of the triangular lattice)
on which the spins are polarized along the $\hat{x}$ direction in response to the transverse field. From the other
two sublattices, it spontaneously chooses one sublattice on which the spins become polarized along the $+\hat{z}$ direction,
while the spins on the third sublattice all point in the $-\hat{z}$ direction in spin space. On the other hand,
when the ordering is of the ferrimagnetic variety, the system spontaneously chooses one sublattice on which the
spins all point along the $+\hat{z}$ direction ($-\hat{z}$ direction), while the spins
on the other two sublattices
all point along the $-\hat{z}$ direction ($+\hat{z}$ direction). 

At low temperature in the presence of an additional next-nearest neighbour ferromagnetic exchange coupling $J_2$ between the $\sigma^z$, one thus expects antiferomagnetic three-sublattice order for small values of $J_2/\Gamma$~\cite{Isakov_Moessner}, and ferrimagnetic three-sublattice order for large
values of $J_2/\Gamma$~\cite{DP_Landau}. The vicinity of this low temperature transition is expected to be a challenging regime for any algorithm, since the ultimate fate of the system
in the vicinity of this transition depends sensitively on the competition
between quantum fluctuations induced by $\Gamma$ and the energetic prerferences
introduced by the presence of a nonzero $J_2$. Nevertheless, we find
that our scheme does not suffer from any significant deterioration in
ergodicity or efficiency in this regime. As a result, we are able to
use it to pinpoint the value of $J_2$ at which the antiferromagnetic three-sublattice order gives way to ferrimagnetic three-sublattice order when we fix $T=0.1$ and $\Gamma=0.8$
in units of $J_1$.
\section{Algorithm}
\label{Method}
We begin with a brief review of the conventional quantum cluster algorithm for transverse field 
Ising models~\cite{Sandvik}, focusing attention on models with a nearest-neighbour
Ising exchange interaction. In this approach, the Hamiltonian is written as a sum of transverse field 
terms living on sites, and Ising exchange interactions living on links of the lattice: 
\begin{equation}
  \begin{split}
    &H_{\mathrm{TFIM}}=-\sum_{a,i}H_{a,i}\\
    & H_{1.i}=|J_{1}|-J_{1}\sigma^{z}_{1(i)}\sigma^{z}_{2(i)}\\
    & H_{2,i}=\Gamma {\mathbf 1}_i\\
    & H_{3,i}=\Gamma \sigma^{x}_{i}
  \end{split}
  \label{Hdecomposition1}
\end{equation}
In the above, the sum over $a$ ranges from $a=1$ to $a=3$.
When $a=1$, the sum over $i$ corresponds to a sum over links of the lattice. On the other hand, when
$a=2$ or $a=3$, the sum  over $i$ corresponds to a sum over sites of the lattice. The composite
labels $1(i)$ and $2(i)$ refer to the two sites 
connected by link $i$, and ${\mathbf 1}_i$ denotes the identity operator acting in the two-dimensional Hilbert space of site $i$. These constant terms $H_{2,i}$ are added to the Hamiltonian to facilitate the quantum cluster updates used in this approach. Additionally,
$H_{1,i}$, which is diagonal in the basis of eigenstates of $\sigma^z$, is shifted
by a constant in order to make all matrix elements positive, which is essential for sign-problem-free Monte Carlo sampling. The operators $H_{a,i}$ acting on any basis state $|\alpha \rangle$ consisting of
a tensor product of eigenstates of $\sigma^z_i$ at various sites $i$ give rise to
a constant multiple of some other basis state $|\beta\rangle$, {\em not a linear
superposition of basis states}:
\begin{equation}
  H_{a,i}|\alpha \rangle \propto |\beta \rangle \; .
\label{nobranch}
\end{equation}
One now expands $Z=\mathrm{Tr}[\exp(-\beta H)]$ 
in a power series in the inverse temperature $\beta$:
\begin{equation}
  Z=\sum^{\infty}_{n=0}\frac{\beta^{n}}{n!}\sum_{S_{n}}\prod^{n}_{k=1}\langle \alpha_{k+1} |H_{a(k),i(k)}|\alpha_{k} \rangle \;,
  \label{SSEeqn}
\end{equation}
where the sum over {\em operator strings} $S_{n}$ runs over all possible sequences of
$n$ operators $H_{a(k),i(k)}$ and basis states $\alpha_k$ ($k=1,2 \dots n$, with the convention that $|\alpha_{n+1}\rangle \equiv |\alpha_1\rangle$). The basic idea now is to perform a Monte Carlo sampling of this sum over operator strings, with
the weight of each operator string set by the product of matrix elements, along with the factor of $\beta^n/n!$. For completeness, we detail below how
this proceeds at an operational level in the conventional algorithm~\cite{Sandvik}.

To change the value of $n$, one uses a {\em diagonal update.} 
To facilitate discussion of this, it is useful to rewrite each term in the expansion Eq.~\eqref{SSEeqn} using a fixed length representation for operator strings. To do this, one
introduces a fixed upper cut-off ${\mathcal L}$ on the length $n$ of operator strings allowed during the simulation, and 
converts each operator string $S_n$ to a length ${\mathcal L}$ object by introducing ${\mathcal L}-n$ identity operators denoted by $H_{0,0}$ in all possible ways. 
There are thus
${\mathcal L} \choose {\mathcal L}-n$ fixed length operator strings corresponding to a single operator string of length $n$. In the fixed-length representation, the analog of Eq.~\eqref{SSEeqn} is thus
\begin{equation}
  Z=\frac{1}{{\mathcal L}!}\sum_{S_{{\mathcal L}}}\beta^{n}({\mathcal L}-n)!\prod^{{\mathcal L}}_{k=1}\langle \alpha_{k+1} |H_{a(k),i(k)}|\alpha_k\rangle 
  \label{LSSEeqn}
\end{equation}
with the convention that $|\alpha_{{\mathcal L}+1}\rangle = |\alpha_1\rangle$.
The diagonal update is now implemented as a sweep through the operator string $S_{{\mathcal L}}$, in which one attempts to replace each diagonal operator by an identity operator $H_{0,0}$ and vice versa. 
The acceptance probabilities are fixed by the requirement that they satisfy the detailed
balance condition with respect to the weight (Eq.~\eqref{LSSEeqn}) of each operator string:
\begin{equation}
  \begin{split}
    & P(H_{0,0}\rightarrow  {\mathrm{diag.}} \; {\mathrm{operator}})=\mathrm{Min}\Big(1,\frac{\beta(N\Gamma+2|J_{1}|N_{links})}{{\mathcal L}-n}\Big),\\
    &  P( {\mathrm{diag.}} \; {\mathrm{operator}} \rightarrow H_{0,0})=\mathrm{Min}\Big(1,\frac{{\mathcal L}-n+1}{\beta(N\Gamma+2N_{\mathrm{links}}|J_{1}|)}\Big),
  \end{split}
  \label{diagup}
\end{equation}
Here, $N$ is the number of lattice sites, and $N_{\mathrm{links}}$ is the number of links in the lattice. On deciding to replace an identity operator by a diagonal operator, the type of diagonal operator is decided such that each 
$H_{2,i}$ is chosen with a probability $\Gamma/(N\Gamma+2|J_{1}|N_{links})$, while each Ising exchange operator $H_{1,i}$ is chosen with a probability
$2|J_{1}|/(N\Gamma+2|J_{1}|N_{links})$. If a particular $H_{2,i}$ is chosen, but not allowed due to  the matrix element corresponding to $H_{2,i}$ in Eq.~\eqref{LSSEeqn} being $0$, then no change
is made at that step of the sweep through the operator string.

\begin{figure}[t]
  \includegraphics[width=7.6cm]{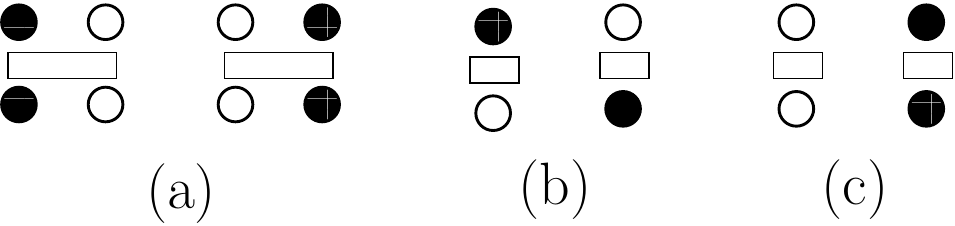}
  \caption{\label{vert1} Allowed vertices from which clusters are built  in the conventional quantum cluster algorithm: (a) Ising exchange operators. (b) Transverse field operators. (c) Constant single site operators. Solid and open circles denote eigenstates of $\sigma^z$ with eigenvalues $+1$ and $-1$ respectively. }
\end{figure}
\begin{figure}[t]
  \includegraphics[width=8.6cm]{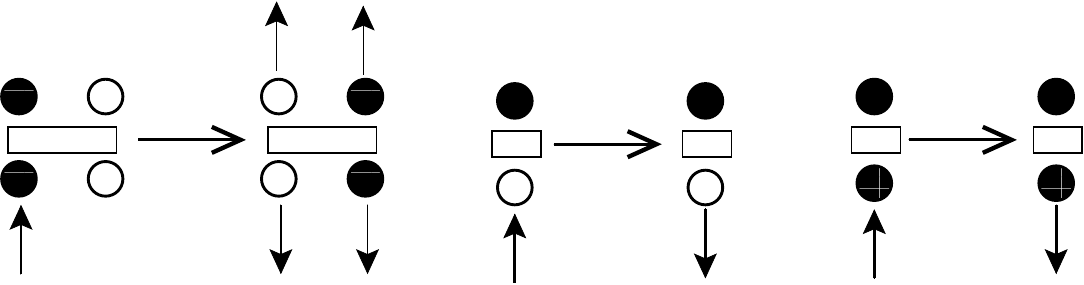}
  \caption{\label{clust1} Cluster construction rules for the conventional quantum
cluster algorithm. Incoming and outgoing arrows show entrance and exit legs respectively.
}
\end{figure}

The other crucial ingredient is the {\em quantum cluster update}. This 
attempts large scale non-local changes in the operator string. To implement this, one maps 
the operator string and the states $|\alpha_k \rangle$ appearing in Eq.~\eqref{LSSEeqn} to a linked vertex list representation. A {\em vertex} corresponds to the non-zero matrix element of a non-identity operator $H_{a,i}$ ({\em i.e.} $a \neq 0$), 
with {\em legs} denoting the spin states of relevant sites in the bra and the ket. The vertices that appear in the construction of clusters in the conventional quantum cluster algorithm
are shown in Fig.~\ref{vert1} for $H_{\mathrm{TFIM}}$ (Eq.~\eqref{Ising}). The Ising-exchange operators correspond to  vertices with four legs each, while the single-site operators
correspond to  vertices with two legs. Links connect a leg of a vertex to corresponding
legs of the next or previous vertex (operator) which acts on that site. In this representation,
the conventional quantum cluster update proceeds as follows: One starts the cluster
construction by choosing a random leg in the operator string. This is the initial {\em entrance leg}. If an entrance leg belongs to an Ising-exchange vertex, all the four legs belonging to this
vertex are added to the cluster. If an entrance leg belongs to a two-leg vertex corresponding to a single-site operator, only that entrance leg is added to the cluster. The cluster building process then proceeds by following the links of all legs added to the cluster (as shown by the outgoing arrows in Fig.~\ref{clust1}). The legs reached by following these links are added to a stack to be processed by the cluster building routine. After this, one repeatedly
takes a new leg
from the stack and checks if it is already part of the cluster being built. If not, this leg
is used as an entrance leg and the same rules are followed. The process ends when
there are no legs left on the stack.
Once a cluster has been made, it can be {\em flipped} by flipping the spin state of all legs that are part of the cluster. Thus, a cluster flip effects operator substitutions of the form $H_{2,i}\rightarrow H_{3,i}$ and  $H_{3,i}\rightarrow H_{2,i}$ . The constant operators $H_{2,i}$ have the same
weight as $H_{3,i}$ in Eq.~\eqref{LSSEeqn}, so flipping a cluster does not change the weight of the operator string.
With this in mind, the conventional quantum cluster algorithm first decomposes the operator string into a set of non-overlapping
clusters using this cluster-building prescription, and then flips each of these clusters with probability $1/2$. One Monte Carlo step of this conventional quantum cluster algorithm is defined as a diagonal update followed by one quantum cluster
update.

To appreciate the challenge faced by this conventional quantum cluster algorithm when
dealing with a frustrated transverse field Ising model, we note that the low temperature
physics in such magnets is dominated by the properties of configurations with minimum Ising exchange energy. In a frustrated Ising model, the number
of such minimally frustrated configurations increases exponentially with the system size,
corresponding to a non-zero $T=0$ limit of the entropy density in the thermodynamic limit
of the classical system.
In the specific case of the triangular lattice antiferromagnet, these minimally frustrated
configurations are characterized by the requirement that there is exactly one
frustrated bond (linking parallel Ising spins) on each triangular plaquette of the lattice.
For $H_{\mathrm{TFIM}}$ on a triangular lattice, the nature of the low-temperature
order is determined by the interplay of thermal and quantum fluctuations acting within this subspace of minimally frustrated configurations~\cite{Moessner_Sondhi,Isakov_Moessner}.

As we will see below, the conventional quantum cluster algorithm presented above fails to efficiently capture the effects of these fluctuations within the subspace of minimally frustrated configurations.
To understand why, we note that the conventional cluster building procedure, being
blind to the distinction between minimally frustrated triangular plaquettes, and fully
frustrated plaquettes (with all Ising spins pointing in the same direction),  either produces very small clusters or branches out along unfrustrated bonds of the lattice to produce a cluster that more often than not spans a very large fraction of the triangular lattice. The small clusters produced by the conventional algorithm accurately capture
the short-distance physics of {\em flippable spins}, which have zero net Ising-exchange
field acting on them, allowing them to flip freely in response to the transverse field $\Gamma$. However, and this is key, the large clusters produced by the conventional algorithm do not capture the important
effects of large-scale fluctuations within the minimally
frustrated subspace, since these large-scale fluctuations do not comprise of simple spin-flips of large parts of the system.

\begin{figure}[t]
  \includegraphics[width=7.6cm]{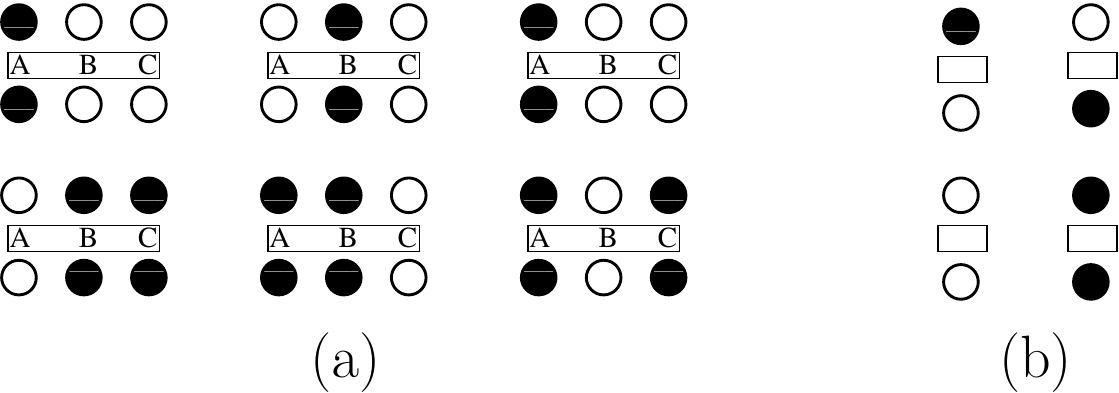}
  \caption{\label{vert2} Allowed vertices in the new quantum cluster algorithm: (a) Triangular plaquette vertices. (b) Single site vertices  (constant operators and transverse field operators).  Solid and open circles denote eigenstates of $\sigma^z$ with eigenvalues $+1$ and $-1$ respectively. }
\end{figure}

In order to overcome these difficulties with the conventional quantum cluster algorithm,
one needs a way of incorporating in our cluster construction
procedure the distinction between minimally frustrated
triangular plaquettes and fully frustrated triangular plaquettes~\cite{KBD,KBD1}.
Following previous work on classical and quantum frustrated systems~\cite{Zhang,Coddington_Han,Melko,Dariush_Kedar,KD_unpublished,gros}, we achieve this by decomposing the antiferromagnetic nearest-neighbour Ising-exchange part of 
$H_{\mathrm{TFIM}}$ as a sum of operators acting on elementary triangular plaquettes. Although this can be done for any frustrated Ising antiferromagnet on a lattice with
triangular plaquettes (such as the triangular, kagome, hyper-kagome or pyrochlore lattice),
 we focus the rest of our discussion on the triangular lattice case for concreteness. In this case, each triangular plaquette has one $A$ sublattice site, one $B$ sublattice
site, and one $C$ sublattice site (here, the labels $A$, $B$, and $C$ 
correspond to the natural tripartite decomposition of the triangular lattice). Thus, the
nearest-neighbour Ising-exchange terms are now written as
\begin{equation}
  H_{1,i}=\frac{3}{2}J_{1}-\frac{J_{1}}{2}\Big(\sigma^{z}_{A(i)}\sigma^{z}_{B(i)}+\sigma^{z}_{B(i)}\sigma^{z}_{C(i)}+\sigma^{z}_{A(i)}\sigma^{z}_{C(i)}\Big)
  \label{Hdecomposition2}
\end{equation}
where $J_{1}>0$ is the nearest neighbour antiferromagnetic exchange coupling, $i$ runs over the elementary triangular plaquettes, and the composite labels $A(i)$, $B(i)$  and $C(i)$ denote the sites belonging to the respective sublattice in the triangular plaquette $i$. The overall factor of half in front of the second term compensates for the fact that each link is shared by two plaquettes, and the first term represents a constant shift
of energy designed to ensure that all matrix elements of $H_{1,i}$ in the $\sigma^z$
basis are positive. The other operators $H_{2,i}$ and $H_{3,i}$ (where $i$ now labels
sites of the lattice) in the decomposition of
the Hamiltonian remain unchanged.  The vertices corresponding to
this decomposition are shown in Fig.~\ref{vert2}.

The diagonal update  proceeds exactly in the same way, 
with the new update probabilities given by
\begin{equation}
  \begin{split}
    & P(H_{0,0}\rightarrow {\mathrm{diag.}} \; {\mathrm{operator}})=\mathrm{Min}\Big(1,\frac{\beta(N\Gamma+2J_{1}N_{\bigtriangleup})}{{\mathcal L}-n}\Big), \\
    &  P({\mathrm{diag.}} \; {\mathrm{operator}} \rightarrow H_{0,0})=\mathrm{Min}\Big(1,\frac{{\mathcal L}-n+1}{\beta(N\Gamma+2N_{\bigtriangleup}J_{1})}\Big),
  \end{split}
  \label{diagup2}
\end{equation}
where $N_{\bigtriangleup}$ denotes the number of elementary triangular plaquettes in the lattice. As before, once the decision to replace an identity operator with a diagonal
operator is made, each diagonal operator $H_{2,i}$ (where $i$ is a site) is chosen with a probability $\Gamma/(N\Gamma+2J_{1}N_{\bigtriangleup})$, while each Ising exchange operator $H_{1,i}$ (where $i$ is a triangular plaquette) is chosen with a probability
$2J_{1}/(N\Gamma+2J_{1}N_{\bigtriangleup})$. If a particular $H_{2,i}$ is chosen, but not allowed due to  the matrix element corresponding to $H_{2,i}$ in Eq.~\eqref{LSSEeqn} being $0$, then no change
is made at that step of the sweep through the operator string.

To design a useful quantum cluster algorithm for such frustrated Ising antiferromagnets, we note that every Ising exchange vertex corresponds to a triangular plaquette on which
two sites host a {\em majority spin} and one site hosts a {\em minority spin}. Thus
each Ising-exchange vertex has two pairs of majority-spin legs and one pair of minority-spin legs. Flipping the spin state of one of the sites that hosts a majority spin corresponds
to flipping the spin state of one pair of majority-spin legs. This  gives another
valid Ising-exchange vertex with the same weight. The new Ising-exchange
vertex obtained in this way satisfies three key properties. First,
this new Ising-exchange vertex {\em does not} correspond to a {\em global spin-flip}
of the {\em entire triangular plaquette}. Second, the pair of majority-spin legs
that were flipped are again majority-spin legs of the new Ising-exchange vertex.
Third, the pair of majority-spin legs that were {\em not flipped} become
minority-spin legs of the new vertex.
The first property suggests that physically important large-scale fluctuations within
the minimally-frustrated subspace are likely to be captured by a cluster algorithm which
assigns a single pair of
majority-spin legs of a vertex to one cluster, while assigning the other two pairs of legs
of the same vertex to a different cluster. The second property suggests
that it should be possible to ensure that such a cluster-construction protocol satisfies
detailed balance so long as the subtlety introduced by the third property is properly
accounted for.

We now build on this intuition to design our new quantum cluster algorithm, valid for any frustrated transverse-field Ising antiferromagnet on
a lattice with triangular plaquettes. In each of the triangular plaquettes $i$ of the lattice, we single out one site as being {\em privileged} and label it ${\mathcal P}_i$. 
For a given lattice, there may be
considerable freedom for the protocol to be followed in choosing these
{\em privileged sites}. If there are $M$ different natural choices for this protocol, one
can either randomly pick one choice $p$ out of these
$M$ at the start of each quantum cluster update step, and use it to obtain the privileged sites $\{ {\mathcal P}_i^p \}$, or cycle through the various choices in some order. 
Using this choice for the privileged site of any triangular
plaquette, we can now mark one pair of {\em privileged legs} and two pairs of {\em ordinary legs} in each Ising-exchange vertex
that lives on this triangular plaquette. Thus all Ising-exchange vertices
that live on a given triangular plaquette are decomposed in the same way into
two privileged legs and four ordinary legs. 

With these preliminaries out of the way, we are now in a position to specify the cluster construction procedure used in our new quantum cluster algorithm:
 If an entrance leg belongs to a two-leg vertex corresponding to a single-site operator, only that entrance leg is added to the cluster. Thus, for two-leg vertices, there is no change from
the procedure followed in the conventional quantum cluster algorithm.
However, if the entrance leg belongs to a six-leg Ising-exchange vertex, the cluster construction rule now depends on whether the privileged legs
of this Ising-exchange vertex are majority-spin legs of this vertex. If the answer is yes,
the rule is as follows: If the entrance leg is a privileged leg, only the
two privileged legs of the vertex are added to the cluster, whereas, if the entrance leg is
an ordinary leg, the four ordinary legs are added to the cluster.
On the other hand, if the privileged legs of this vertex are {\em not} majority-spin legs, then
all six legs of this vertex are added to the cluster regardless of the entrance leg.
\begin{figure}[t]
  \includegraphics[width=8.6cm]{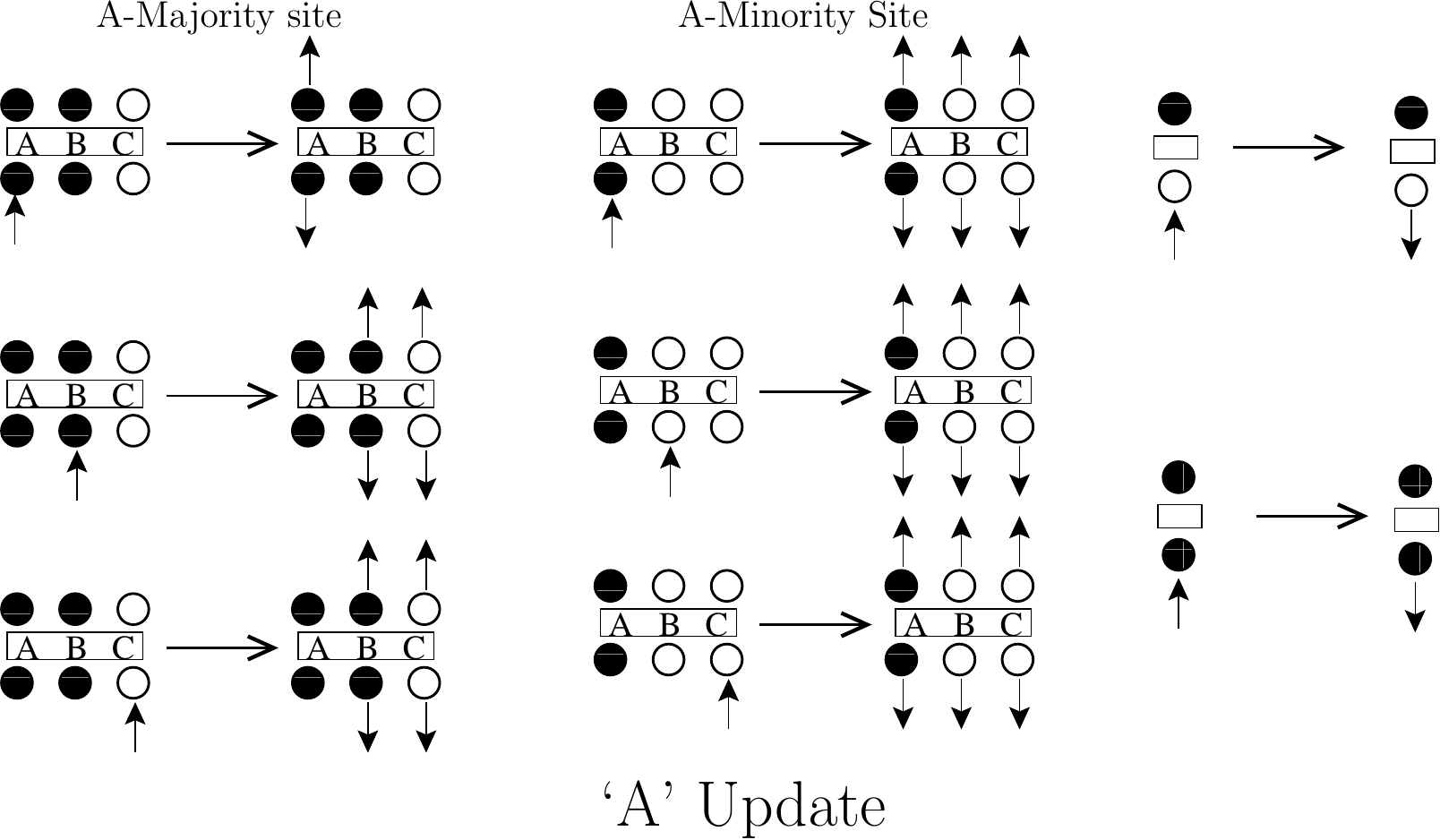}
  \caption{\label{clust2}Cluster construction rules for the `A' update, shown for representative vertices .  The 'B' and 'C' updates proceed in similar ways.
  Incoming and outgoing arrows denote entrance and exit legs respectively.
}
\end{figure}

Specializing this to the case of the triangular lattice Ising antiferromagnet,
we note that there are three natural choices of protocol for choosing the privileged sites,
corresponding to the natural tripartite decomposition of the triangular lattice
into $A$, $B$ and $C$ sublattices. Thus, the privileged site of each
triangular plaquette can be consistently chosen to be the $A$-sublattice site,
or the $B$-sublattice site, or the $C$-sublattice site, and the procedure is
to randomly pick one out of these three choices at the start of each cluster update
step, giving us either a {\em A-update} or a {\em B-update} or a {\em C-update}.
The rules for the {\em A-update} are depicted pictorially in Fig.~\ref{clust2}.
In each update, we construct all possible clusters and then flip each with a probability $1/2$.  As in the conventional quantum cluster algorithm, one Monte Carlo step can be defined as a set of diagonal updates and quantum cluster
updates. In practice, we choose to define a Monte Carlo step
as one cycle through a set of A, B, and C quantum cluster updates interspersed with diagonal updates.

Finally, we sketch the incorporation of additional exchange couplings, using
the triangular lattice transverse-field Ising antiferromagnet as a paradigmatic example.
Consider first the case with an additional {\em ferromagnetic} next-nearest neighbour coupling $J_2<0$. In this case, one
has additional diagonal operators in the decomposition of the Hamiltonian:
\begin{equation}
  \begin{split}
    &H_{\mathrm{TFIM}}=-\sum_{a,i}H_{a,i}\\
    & H_{1,i}=\frac{3}{2}J_{1}-\frac{J_{1}}{2}\Big(\sigma^{z}_{A(i)}\sigma^{z}_{B(i)}+\sigma^{z}_{B(i)}\sigma^{z}_{C(i)}+\sigma^{z}_{A(i)}\sigma^{z}_{C(i)}\Big)\\
    & H_{2,i}=\Gamma {\mathbf 1}_i\\
    & H_{3,i}=\Gamma \sigma^{x}_{i}\\
   & H_{4,i}=|J_{2}|-J_{2}\sigma^{z}_{1(i)}\sigma^{z}_{2(i)}
  \end{split}
  \label{Hdecomposition3}
\end{equation}
In the above, the sum over $a$ ranges from $a=1$ to $a=4$.
When $a=1$, the sum over $i$ corresponds to a sum over triangular plaquettes of the lattice as discussed in the foregoing. When $a$ is $2$ or $3$, the sum  over $i$ corresponds to a sum over sites of the lattice as before. Finally, when $a=4$, the sum over
$i$ is over links of the lattice, since we use the original link-based decomposition
for ferromagnetic interactions. The composite
labels $1(i)$ and $2(i)$ in this case refer to the two sites 
connected by link $i$. Thus, the operator string now has six-leg vertices corresponding
to antiferromagnetic nearest-neighbour Ising-exchange couplings, four-leg vertices
corresponding to ferromagnetic next-nearest-neighbour Ising-exchange couplings,
and two-leg vertices that represent the single-site operators.
The quantum cluster update now uses our new procedure for handling
six-leg vertices while continuing to use the original rules of the conventional
quantum cluster update when dealing with four-leg vertices. The diagonal
update also involves obvious changes, since there are now three kinds
of diagonal operators.

On the other hand, if the next-nearest neighbour coupling is {\em antiferromagnetic} ($J_2 >0$), we use a triangular plaquette based decomposition of the new terms in the
Hamiltonian, with the new triangular plaquettes being the elementary
triangular  plaquettes of the three triangular lattices formed from the $A$, $B$
and $C$ sublattice sites of the original triangular lattice. Thus, the
operator string now has two different kinds of six-leg vertices, apart
from the two-leg vertices that represent single-site operators.

The modifications needed to account for the different kinds of diagonal vertices
in the diagonal update are again straightforward, and not spelled out here. The only
subtlety is that our cluster construction rules must now use two different
notions of privileged sites when dealing with
the two different kinds of six-leg vertices. For the six-leg vertices
corresponding to the nearest neighbour antiferromagnetic interactions, we
choose the privileged sites to be either the $A$, $B$ or $C$ sublattice sites of the triangular lattice.
For the six-leg vertices corresponding to next-nearest neighbour antiferromagnetic
interactions, one can think of several different notions of privileged sites. The one
we have tested most thoroughly in our work uses a three-sublattice decomposition
of the three new triangular lattices to define the privileged sites for the next-nearest
neighbour interaction vertices. This leads to $3^4 \equiv 81$ different possibilities
for the full cluster construction rules, each of which obeys detailed balance.
In our simulations, we define a Monte Carlo step
as one cycle through a set of three randomly chosen quantum cluster updates interspersed with diagonal updates.
\section{Results}
\label{Results}
Our simulations use both the conventional and the new quantum cluster algorithm
to study the transverse field Ising antiferromagnet on $L\times L$ triangular lattices with periodic boundary conditions and $L$ a multiple of six ranging from $L=36$ to $L=108$.
To facilitate direct comparison of the two algorithms, we define one Monte Carlo
step of the conventional quantum cluster algorithm in the same way as the
new quantum cluster algorithm, {\em i.e.} as a set of three cluster updates
interspersed with diagonal updates.

A transverse field $\Gamma$ is known to induce long-range three-sublattice order
in the Ising antiferromagnet on a triangular lattice~\cite{Isakov_Moessner}. This order melts in a two-step manner through an intermediate phase with power-law three-sublattice
order~\cite{Isakov_Moessner}. Long-range three-sublattice order is known
to persist up to the highest temperatures in the vicinity of an optimal
value of the transverse field of around $\Gamma \approx 0.8$~\cite{Isakov_Moessner} in
units of $J_1$. Therefore, we choose $J_1=1$ and a transverse field of $\Gamma=0.8$ 
in many of the simulations performed to compare the performance of the new algorithm
with that of the conventional quantum cluster algorithm. Most
of our results are deep in the ordered state, which extends to roughly
a temperature of $T \approx 0.2$ in units of $J_1$ when $\Gamma = 0.8$~\cite{Isakov_Moessner}.

We study the physics of three-sublattice ordering in such magnets by
computing the  the complex three-sublattice order parameter $\psi$ and the kinetic energy density at the three-sublattice ordering wave vector $\epsilon_{\mathbf{Q}}$. Additionally,
we also measure  the uniform easy axis magnetization $m$. These quantities are
defined as
\begin{eqnarray}
m&=&\frac{1}{L^{2}}\sum_{\vec{r}}\sigma^{z}_{\vec{r}} \\
\psi&=&\frac{1}{L^{2}}\sum_{r}\sigma^{z}_{\vec{r}}\exp(i\mathbf{Q} \cdot \vec{r}) \\
\epsilon_{\mathbf{Q}}&=&\frac{1}{L^{2}}\sum_{r}\sigma^{x}_{\vec{r}}\exp(i\mathbf{Q} \cdot \vec{r})
\end{eqnarray}
where $\mathbf{Q}$
is the three-sublattice ordering wave vector and $\vec{r}$ represents the coordinates
of triangular lattice sites.
The corresponding static susceptibilities are defined in the standard way
\begin{eqnarray}
\chi_{0}&=&\frac{L^{2}}{\beta}\langle\lvert \int_{0}^{\beta} d\tau  m(\tau)\rvert^{2}\rangle\\
\chi_{{\mathbf{Q}}}&=&\frac{L^{2}}{\beta}\langle\lvert \int_{0}^{\beta} d\tau \psi(\tau)\rvert^{2}\rangle\\
\chi^{xx}_{{\mathbf{Q}}}&=&\frac{L^{2}}{\beta}\langle\lvert\int_{0}^{\beta} d\tau\epsilon_{\mathbf{Q}}(\tau)\rvert^{2}\rangle
\end{eqnarray}
where 
${\mathcal O}(\tau) \equiv e^{\tau H_{\rm TFIM}}{\mathcal O}e^{-\tau H_{\rm TFIM}}$ denotes the conventional imaginary-time version of the Heisenberg operator corresponding
to any Schrodinger operator ${\mathcal O}$. Monte Carlo estimators for
these quantities, presented in Ref.~\cite{SandvikSSE}, can be used without any changes
with the new quantum cluster algorithm.

\begin{figure}[t]
  \includegraphics[width=8.6cm]{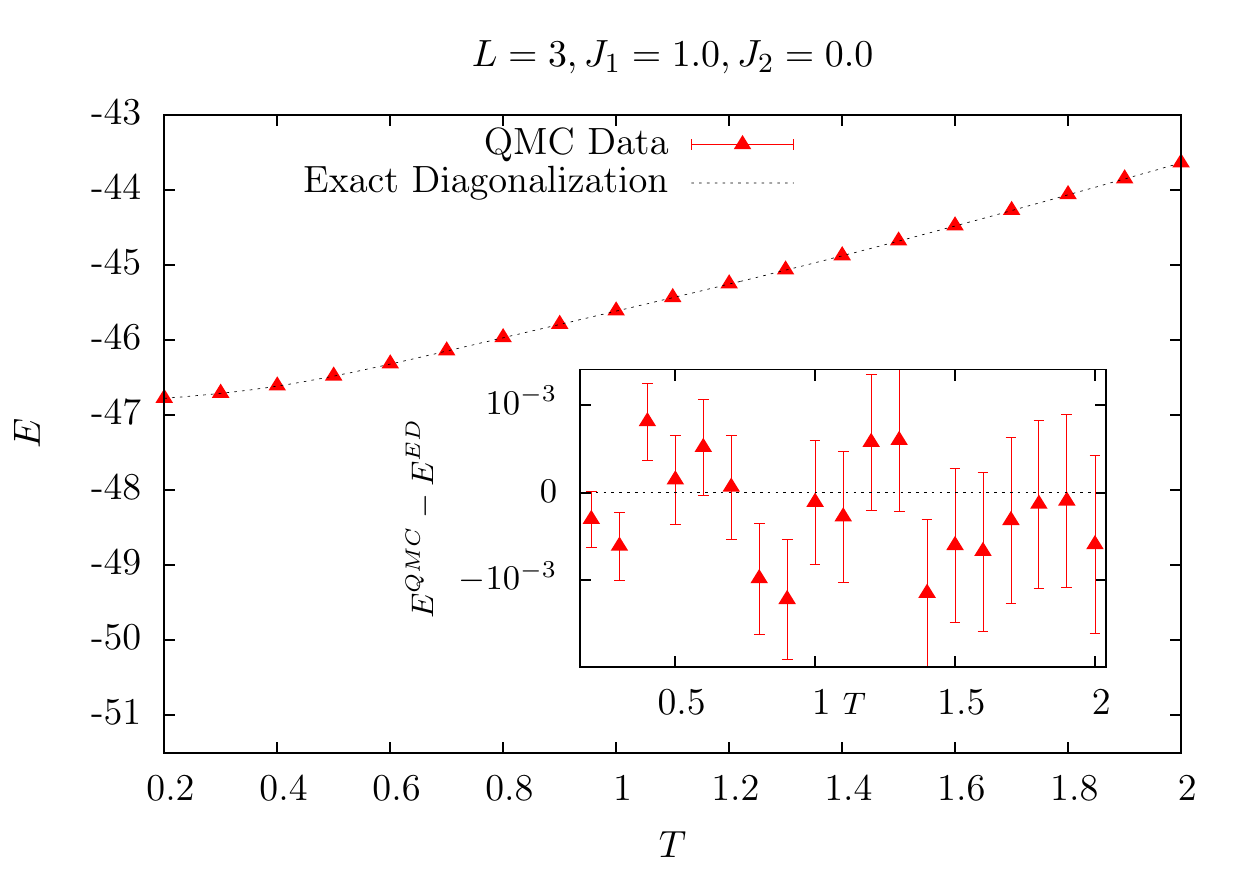}
  \caption{\label{EDe} Quantum Monte Carlo data (obtained using the new quantum
cluster algorithm) for the  total energy $E$ of a triangular lattice transverse field Ising antiferromagnet on a $3 \times 3$ lattice compared with exact diagonalization results for the same quantity. Inset zooms in on the statistical fluctuations of the
Monte Carlo results.}
\end{figure}

\begin{figure}[t]
  \includegraphics[width=8.6cm]{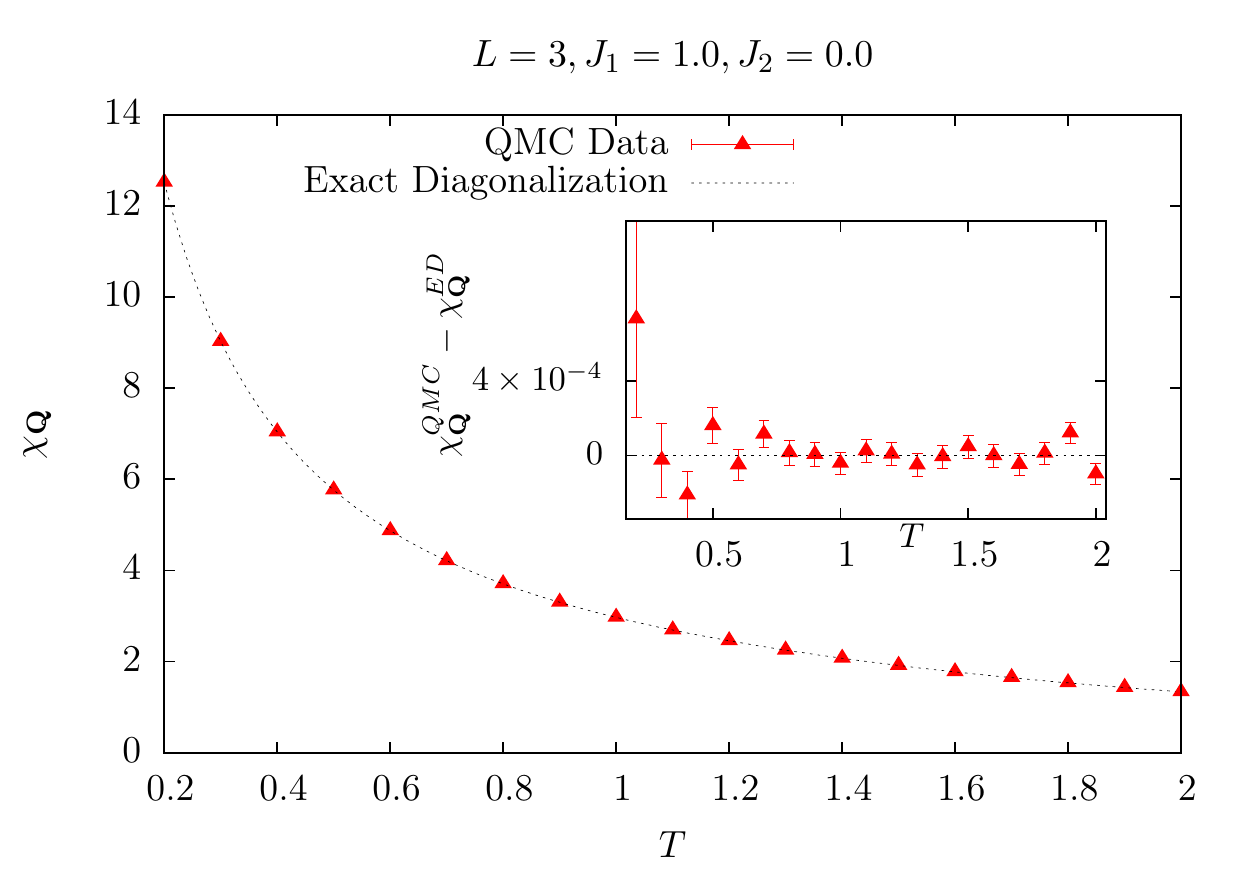}
  \caption{\label{EDchi}Quantum Monte Carlo data (obtained using the new quantum
cluster algorithm) for the spin susceptibility $\chi_{\mathbf{Q}}$ of a triangular lattice transverse field Ising antiferromagnet on a $3 \times 3$ lattice compared with exact diagonalization results for the same quantity. Inset zooms in on the statistical fluctuations of the
Monte Carlo results.}
\end{figure}

\begin{figure}[t]
  \includegraphics[width=8.6cm]{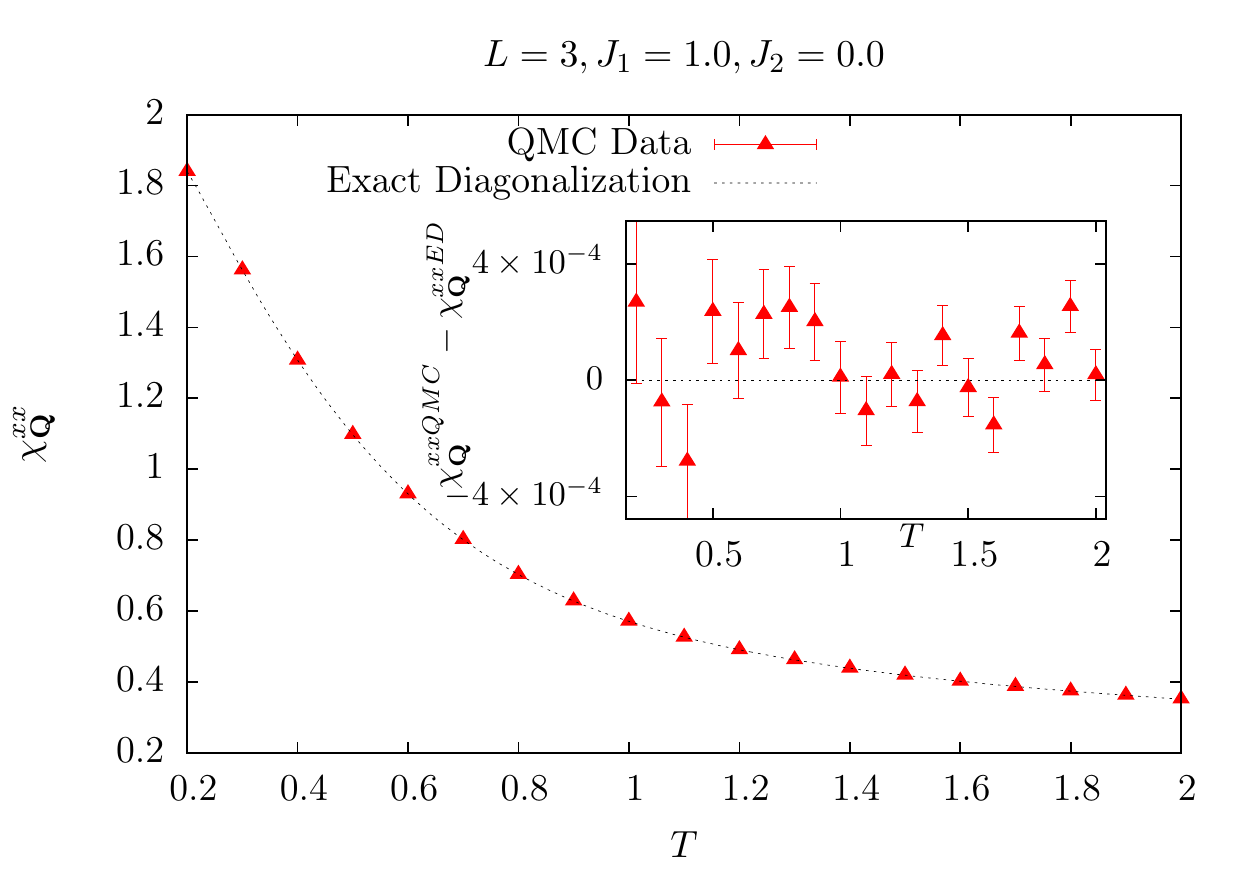}
  \caption{\label{EDchix}Quantum Monte Carlo data (obtained using the new quantum
cluster algorithm) for the transverse spin susceptibility $\chi^{xx}_{\mathbf{Q}}$ of a triangular lattice transverse field Ising antiferromagnet on a $3 \times 3$ lattice compared with exact diagonalization results for the same quantity. Inset zooms in on the statistical fluctuations of the
Monte Carlo results.}
\end{figure}

To test our implementation of the new quantum cluster algorithm, we have first 
obtained high precision results for a small system of linear size $L=3$
and compared these against the results of exact diagonalization. Results
of such tests, for a simulation with $2\times10^{8}$ Monte Carlo steps  preceeded by a warm-up of $2\times10^{8}$ Monte Carlo steps, is shown in Fig.~\ref{EDe} for the total energy $E$,
and in Fig.~\ref{EDchi} and Fig.~\ref{EDchix} for $\chi_{\mathbf{Q}}$  and $\chi^{xx}_{\mathbf{Q}}$ respectively.
\begin{figure}[t]
  \includegraphics[width=8.6cm]{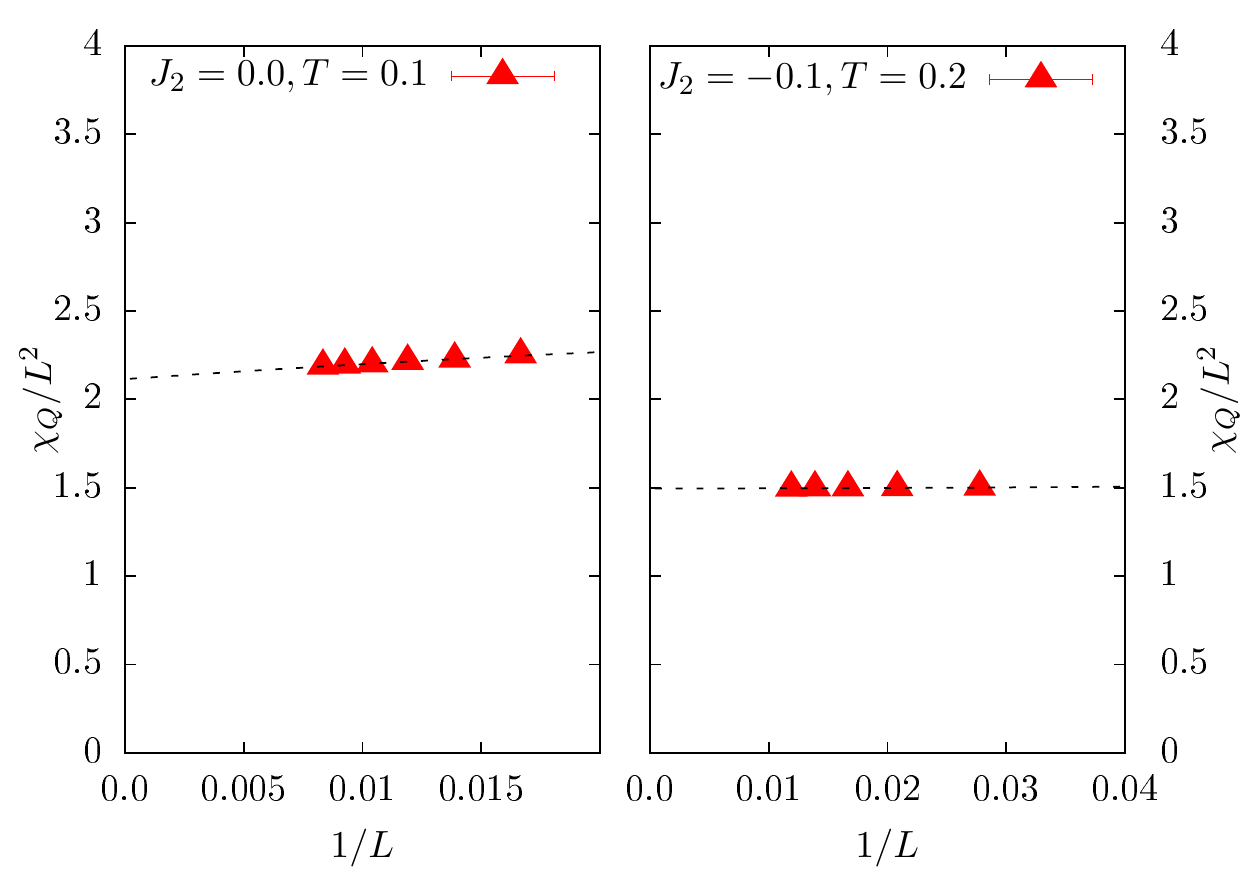}
  \caption{\label{ordered} $H_{\rm TFIM}$ is three-sublattice ordered both at $J_{2}=0.0$, $T=0.1$ (left panel) and at $J_{2}= -0.1$, $T=0.2$ (right panel) when $J_1=1$. This is established by confirming that $\chi_{Q}/L^{2}$ extrapolates to a non-zero value as $1/L \rightarrow \infty$. Lines are best fits to the form $a+b/L+c/L^{2}$.}
\end{figure}

Since the physics of the three-sublattice ordered low temperature state is dominated by the effects of
quantum and thermal fluctuations within the subspace of minimally frustrated classical
configurations, this represents the most challenging regime for the Monte Carlo work.
With this in mind, we focus on a comparison of the two algorithms deep in
this ordered state. In Fig.~\ref{ordered}, we first confirm that $H_{\rm TFIM}$ does
indeed have long-range three-sublattice order at $T=0.1$ when $J_1=1$~\cite{Isakov_Moessner} and
there is no next-nearest-neighbour coupling $J_2$.
In this ordered state, we characterize the performance of the two algorithms by measuring the autocorrelation function of the Monte Carlo
estimators of the physical observables defined in the foregoing. 
 
\begin{figure}[t]
  \includegraphics[width=8.6cm]{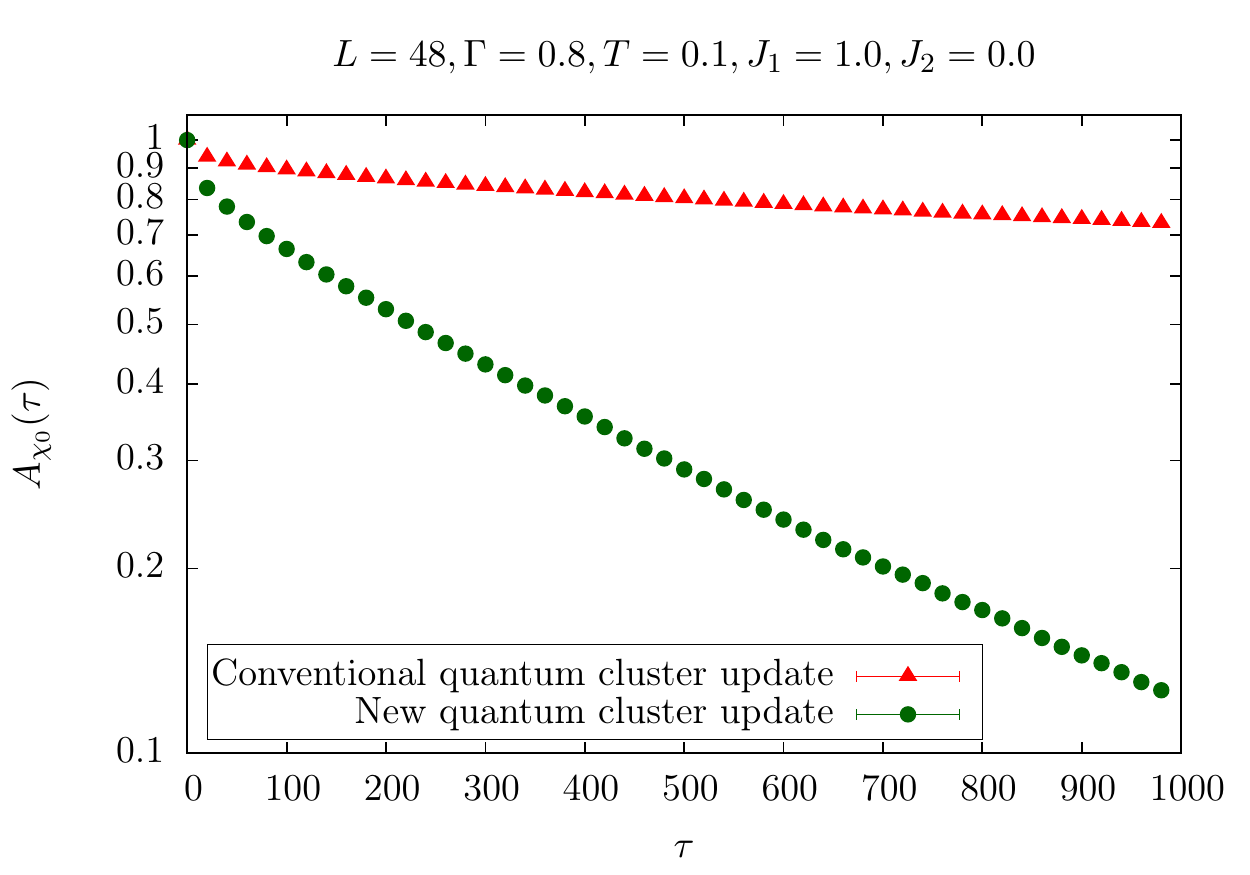}
  \caption{\label{chi048} Auto-correlation function of the Monte Carlo estimator
of $\chi_{0}$ for the triangular lattice transverse field Ising antiferromagnet on
an $L \times L$ lattice with $L=48$. The x-axis label $\tau$ refers to the number
of Monte Carlo steps.}
\end{figure}

\begin{figure}[t]
  \includegraphics[width=8.6cm]{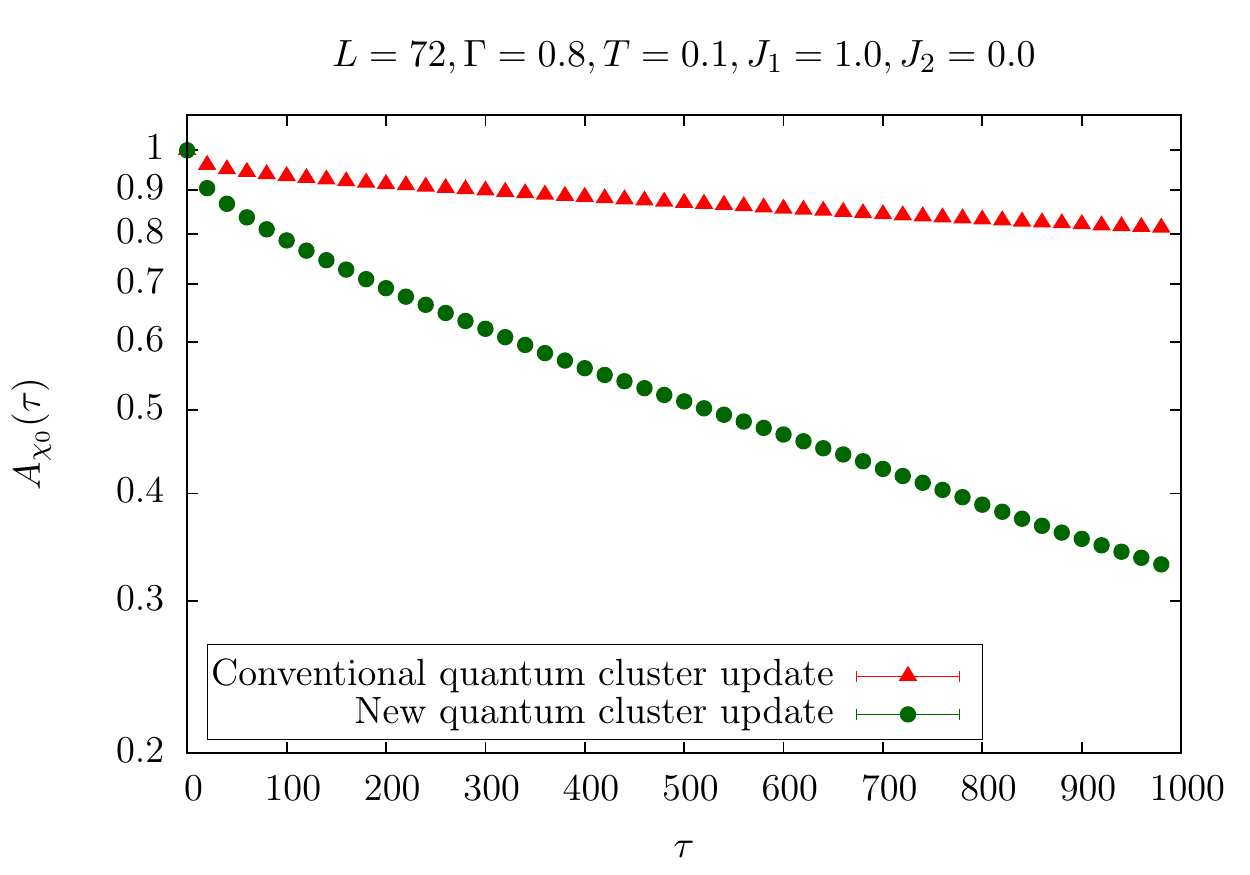}
  \caption{\label{chi072} Auto-correlation function of the Monte Carlo estimator
of $\chi_{0}$ for the triangular lattice transverse field Ising antiferromagnet on
an $L \times L$ lattice with $L=72$. The x-axis label $\tau$ refers to the number
of Monte Carlo steps.}
\end{figure}

\begin{figure}[t]
  \includegraphics[width=8.6cm]{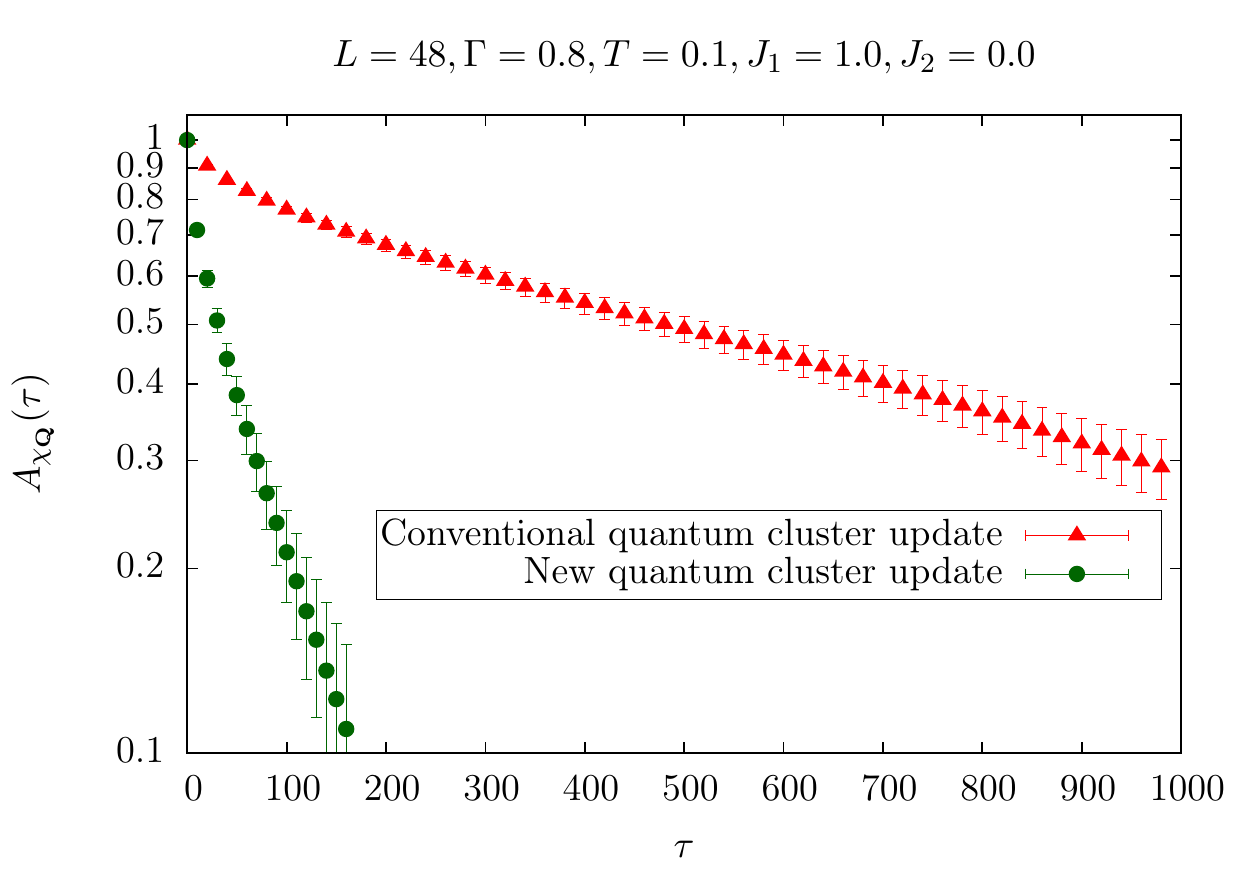}
  \caption{\label{chiQ48}Auto-correlation function of the Monte Carlo estimator
of $\chi_{{\mathbf{Q}}}$ for the triangular lattice transverse field Ising antiferromagnet on
an $L \times L$ lattice with $L=48$. The x-axis label $\tau$ refers to the number
of Monte Carlo steps.}
\end{figure}

\begin{figure}[t]
  \includegraphics[width=8.6cm]{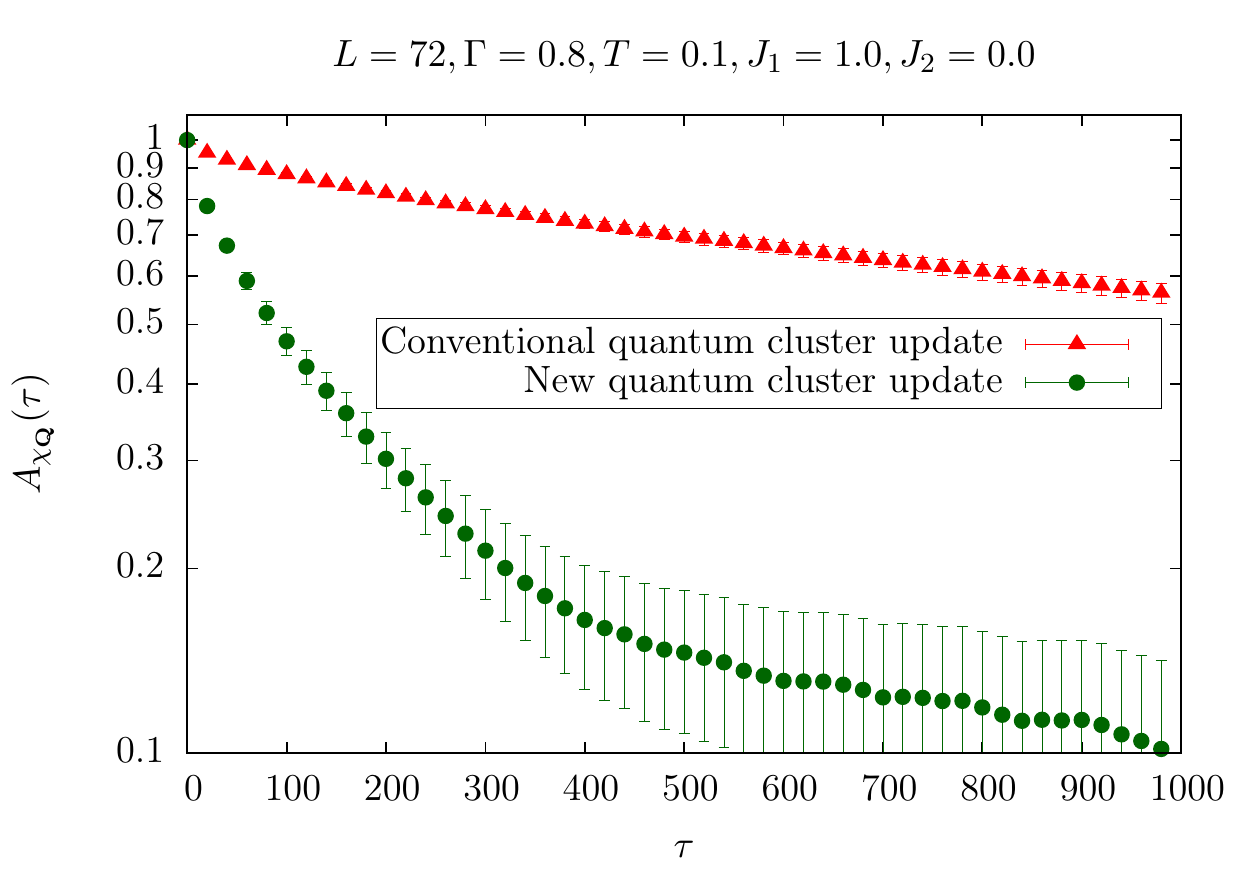}
  \caption{\label{chiQ72}Auto-correlation function of the Monte Carlo estimator
of $\chi_{{\mathbf{Q}}}$ for the triangular lattice transverse field Ising antiferromagnet on
an $L \times L$ lattice with $L=72$. The x-axis label $\tau$ refers to the number
of Monte Carlo steps.}
\end{figure}

For an observable $O$, with Monte Carlo estimator given by $O(\tau)$, the autocorrelation function  $A_{O}(\tau)$, for a run of $\tau_{\mathrm{m}}$ Monte Carlo steps,  is defined as
\begin{equation}
  A_{O}(\tau)=\langle O(n+\tau)O(n)\rangle -\Big(\langle O(n)\rangle \Big)^{2} \; ,
\end{equation}
where the angular brackets denote averaging of $O(n)$ over the Monte Carlo configurations generated in the simulation. After obtaining this function,
we normalize it so that $A_{O}(0)=1$. The function $A_{\chi_{0}}$, normalized in
this way, is displayed in Fig.~\ref{chi048} and Fig.~\ref{chi072}  for systems of linear size $L=48$ and $L=72$ respectively. The corresponding function $A_{\chi_{\mathbf{Q}}}$, normalized
in the same way, is displayed in Fig.~\ref{chiQ48} and Fig.~\ref{chiQ72} for the same
values of $L$.  All these autocorrelation
measurements are performed during a Monte Carlo simulation
of $2\times10^{6}$ Monte Carlo steps, after a warm up of $2\times10^{5}$ Monte Carlo steps.

\begin{figure}[]
  \includegraphics[width=8.6cm]{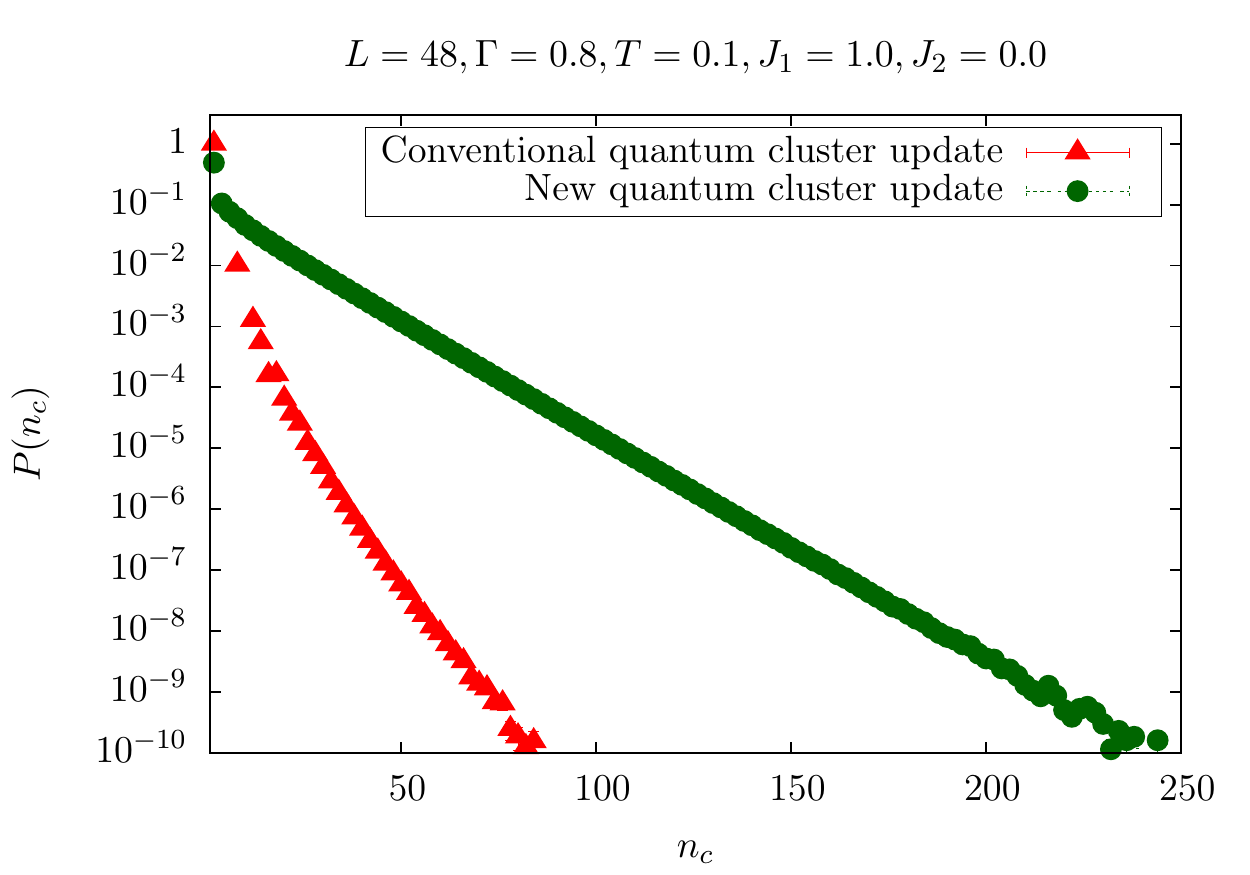}
  \caption{\label{small}The distribution of small clusters built by the conventional
quantum cluster algorithm and the new quantum cluster algorithm during simulations
of the triangular lattice transverse field Ising antiferromagnet, displayed as a function of $n_{c}$, the cluster size (number of legs that belong to a cluster).}
\end{figure}

We note that the auto-correlation function for both algorithms are not simple exponentials
in the ordered state at $J_2=0$.
Rather, they appear to have multiple time-scales.  Further, it is clear that the new quantum cluster algorithm is significantly more efficient at decorrelating the Monte Carlo
estimators of quantities studied, and this advantage appears to be
roughly independent of size.
To investigate the mechanics behind this fairly dramatic improvement at low temperature, we take the number of vertex-legs $n_c$ in a cluster as a measure of cluster size, and look at the statistics of sizes of clusters constructed in the ordered state, as well as at
higher temperatures. 
Our first observation is that both algorithms make two kinds of clusters, {\em small}
and {\em large}. Small clusters have a size that is roughly independent of $n_{\rm tot}$, the
total number of legs in the operator string,
while large clusters have a size that scales with $n_{\rm tot}$. We find
that the distribution of small clusters
does not depend much on system size and temperature for either algorithm, being presumably set by the short-distance physics of {\em flippable} spins and clusters of spins,
which have no net exchange field on them due to interactions with the rest of the system.
When compared to the distribution of small clusters made by the conventional
quantum cluster algorithm, the corresponding distribution for the new quantum
cluster algorithm is nevertheless considerably broader, as is clear from Fig.~\ref{small}.
We ascribe part of the improved performace of the new quantum cluster
algorithm to this difference in the distributions of small clusters.
\begin{figure}[]
  \includegraphics[width=8.6cm]{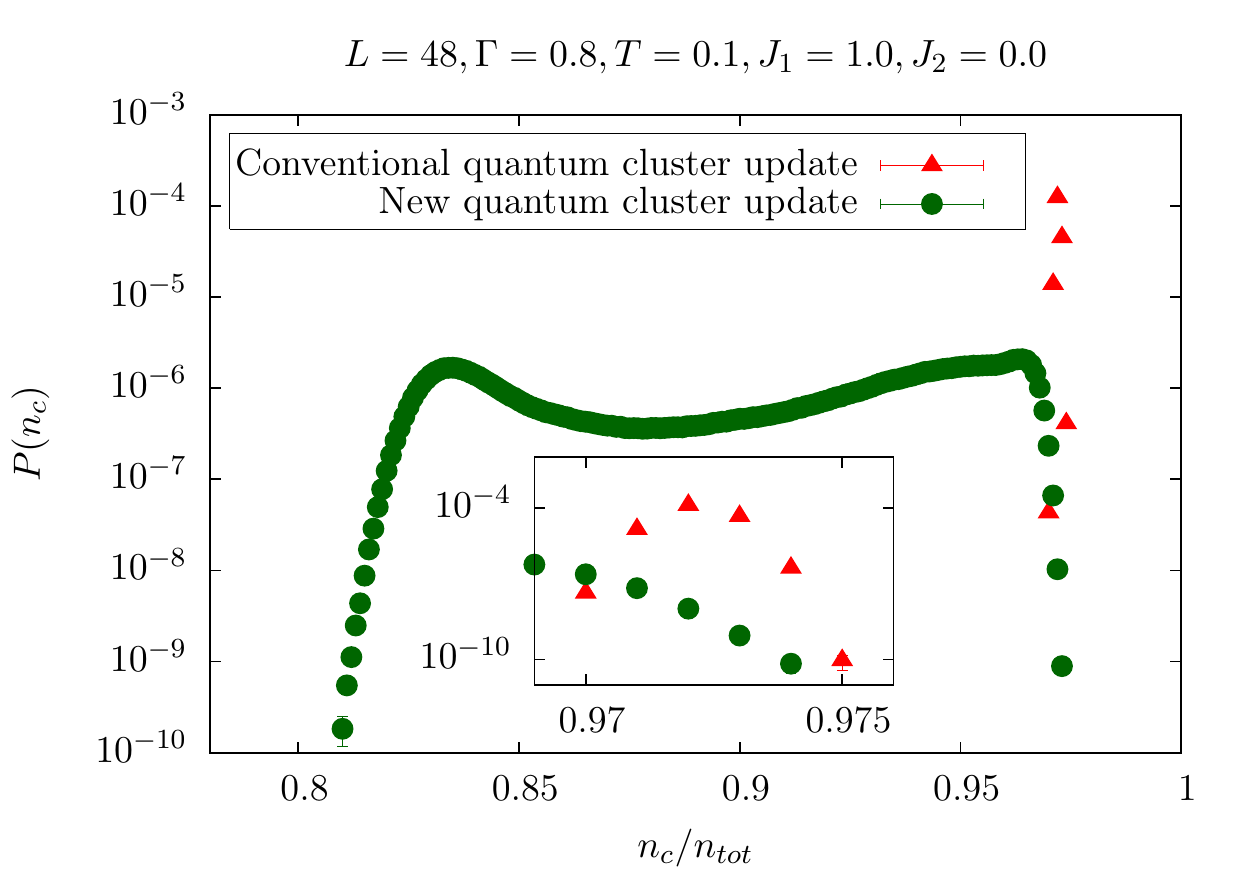}
  \caption{\label{large}The distribution of large clusters built by the conventional
quantum cluster algorithm and the new quantum cluster algorithm during simulations
of the triangular lattice transverse field Ising antiferromagnet, displayed as a function of $n_{c}/n_{\rm tot}$, the cluster size (number of legs that belong to a cluster) normalized by
the total number of legs in the operator string. Inset zooms in on
the extremely narrow peak that dominates this distribution for the conventional
quantum cluster algorithm.}
\end{figure}
\begin{figure}[]
  \includegraphics[width=8.6cm]{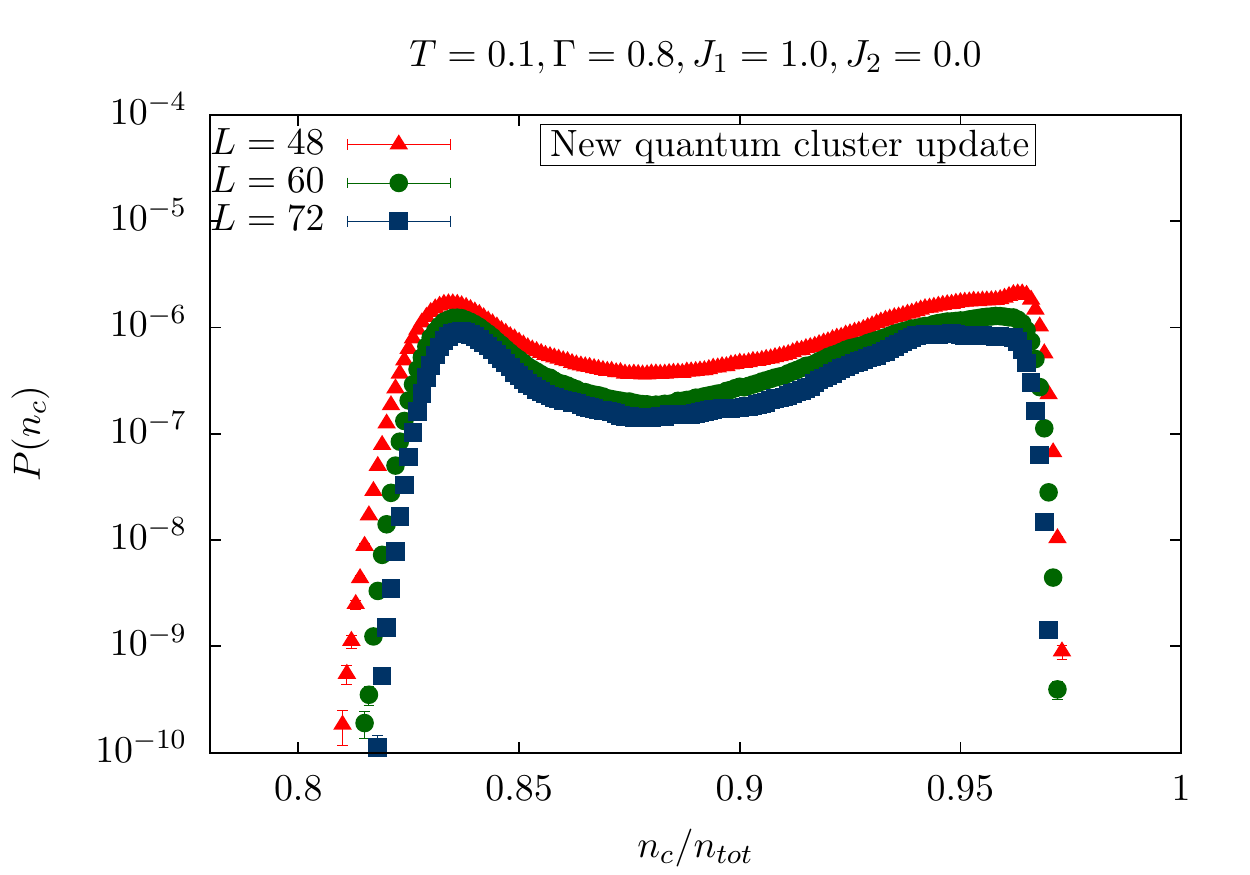}
  \caption{\label{distsize} System size dependence of the distribution of large clusters built by the new quantum cluster algorithm during simulations of the triangular lattice transverse field Ising antiferromagnet, displayed as a function of $n_{c}/n_{\rm tot}$, the cluster size (number of legs that belong to a cluster) normalized by
the total number of legs in the operator string. }
\end{figure}
\begin{figure}[]
  \includegraphics[width=8.6cm]{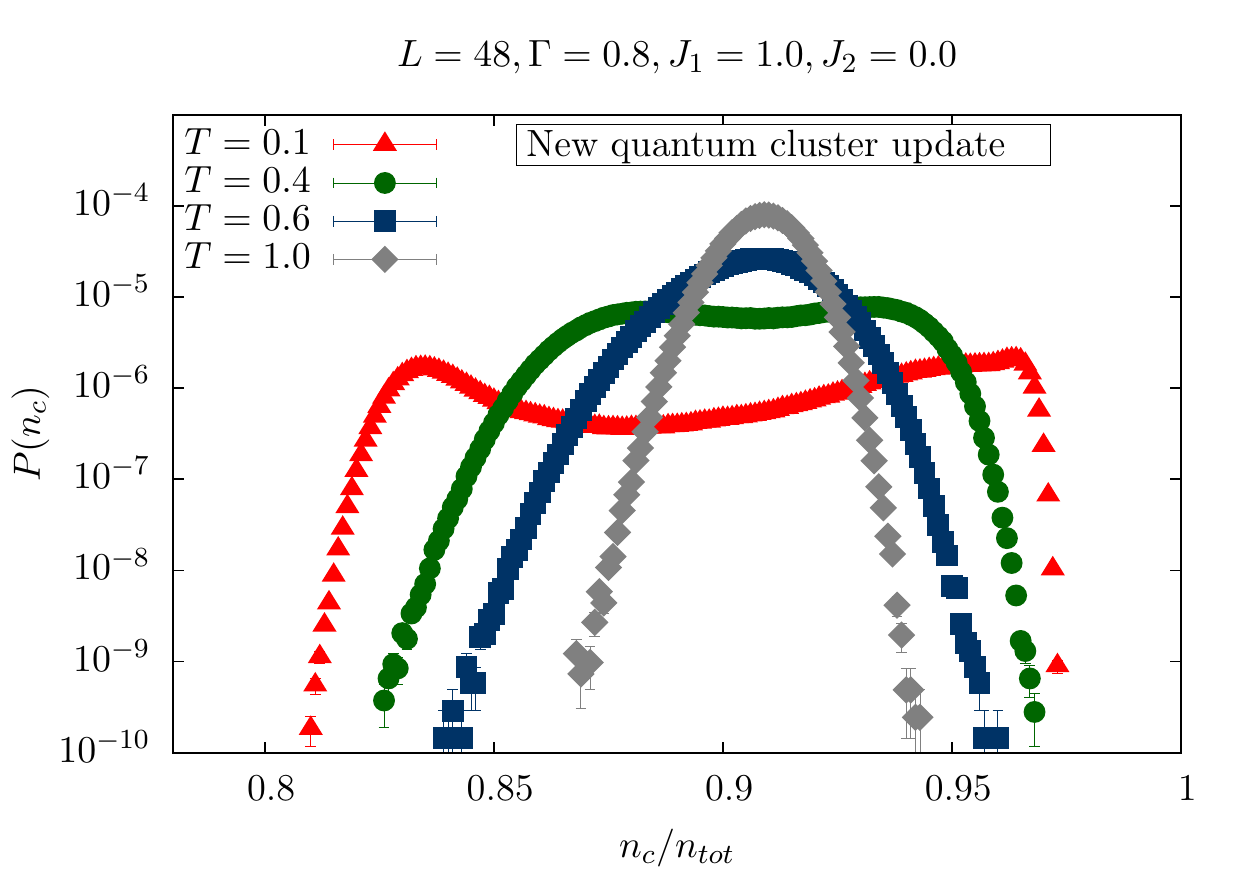}
  \caption{\label{disttemp}Temperature dependence of the distribution of large clusters built by the new quantum cluster algorithm during simulations of the triangular lattice transverse field Ising antiferromagnet, displayed as a function of $n_{c}/n_{\rm tot}$, the cluster size (number of legs that belong to a cluster) normalized by
the total number of legs in the operator string. }
\end{figure}

The other reason for the dramatically better performance of the new quantum cluster
algorithm is clear from a comparison of the distribution of large clusters made by the two
algorithms in the ordered state. This comparison is shown in Fig.~\ref{large}.
Clearly, the conventional quantum cluster algorithm mostly makes clusters that,
when flipped, amount to nearly a global spin flip of the entire operator string, while
the new quantum cluster algorithm produces a fairly broad distribution of
$n_c/n_{\rm tot}$ which presumably plays a major role in improving the
performance of the algorithm.
For our new algorithm, we also monitor the change
in this distribution of large clusters
as a function of system size (Fig.~\ref{distsize}) and temperature (Fig.~\ref{disttemp}).
As is clear from these figures, the distribution remains usefully broad even at larger sizes at low
temperature, and narrows somewhat only on exiting the ordered state upon
heating.

\begin{figure}[t]
  \includegraphics[width=8.6cm]{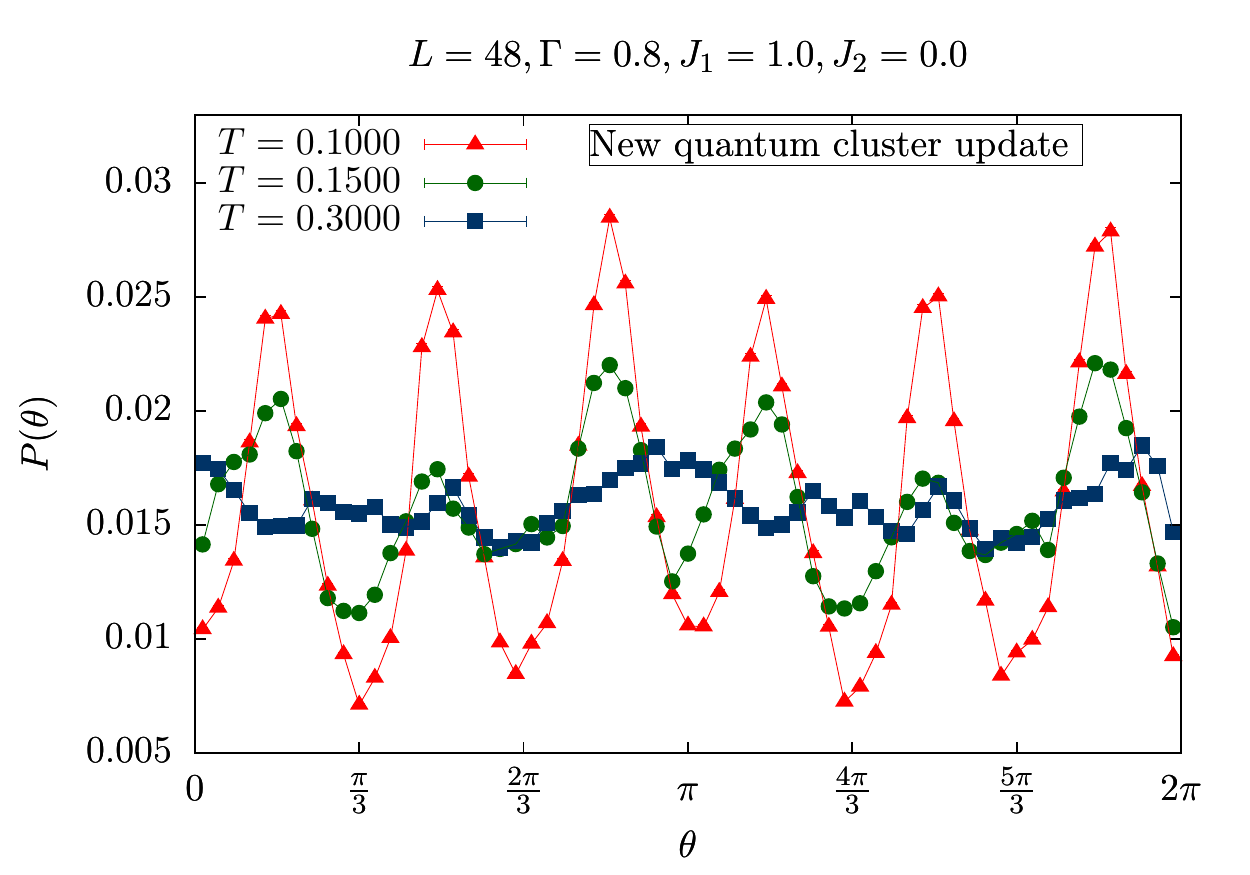}
  \caption{\label{48PAF}Histograms of the phase of the estimator $\theta$ of the
complex three-sublattice order parameter $\psi$, obtained during
simulations performed using the new quantum cluster algorithm with $J_2=0$,
$J_1=1$ and $\Gamma=0.8$. Note the
six peaks at $(2m+1)\pi/6$ ($m=1.2 \dots 6$) in the low temperature state, characteristic of antiferromagnetic
three-sublattice order.}
\end{figure}

\begin{figure}[t]
  \includegraphics[width=8.6cm]{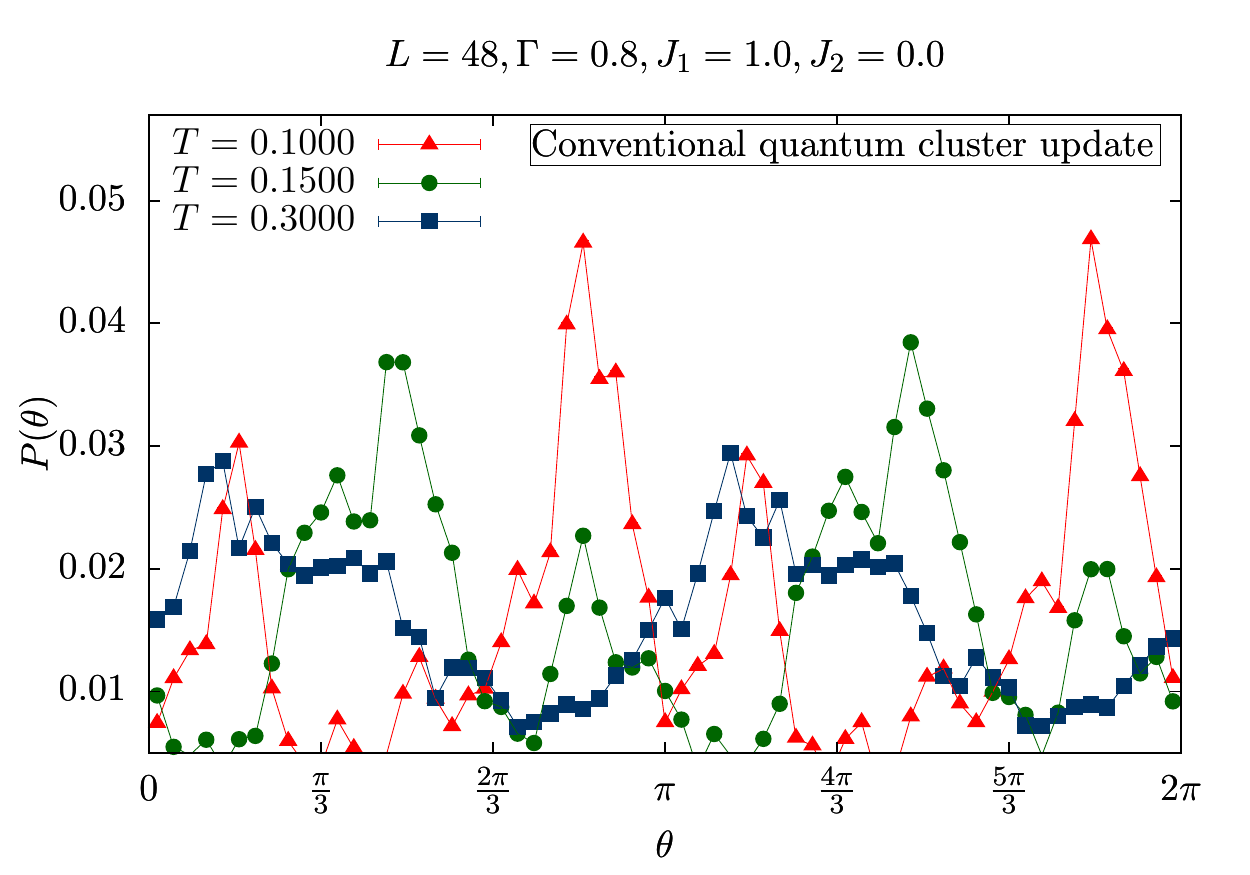}
  \caption{\label{48LAF}Histograms of the phase of the estimator $\theta$ of the
complex three-sublattice order parameter $\psi$, obtained during
simulations performed using the conventional quantum cluster algorithm with $J_2=0$,
$J_1=1$ and $\Gamma=0.8$. Note that
it is hard to distinguish six clearly demarcated peaks at $(2m+1)\pi/6$ ($m=1.2 \dots 6$) in the low temperature state, which
would have been indicative of antiferromagnetic three-sublattice order.}
\end{figure}

Histograms of the phase $\theta$ of the Monte Carlo estimator for the
complex order parameter $\psi$
provide another fairly striking signature of the improved performance of
the new quantum cluster algorithm. For instance, deep in the ordered
state with $J_2=0$, the histogram of $\theta$ produced by the new
quantum cluster algorithm exhibits six well-defined and clearly separated peaks
at the values $(2m+1)\pi/6$ ($m=1,2 \dots 6$), characteristic of antiferromagnetic three
sublattice order~\cite{KDamle_PRL}. Upon heating, the peaks
get less pronounced, and finally disappear in the power-law ordered
phase, providing a nice visual representation of the fact
that the long-wavelength physics of the power-law ordered phase is expected to be controlled by a $U(1)$ symmetric fixed line, with
six-fold anisotropy in $\theta$ expected to
be an irrelevant perturbation of this fixed line~\cite{Jose_Kadanoff,KDamle_PRL}. This is shown in Fig.~\ref{48PAF}. In sharp contrast, the corresponding histograms obtained from the conventional quantum cluster algorithm are not as conclusive. In the ordered state, it is not possible
to clearly distinguish six peaks at the values $(2m+1)\pi/6$, nor is there a clean
distinction between the power-law ordered regime and the low-temperature ordered phase.
This is shown in Fig.~\ref{48LAF}.

\begin{figure}[t]
  \includegraphics[width=8.6cm]{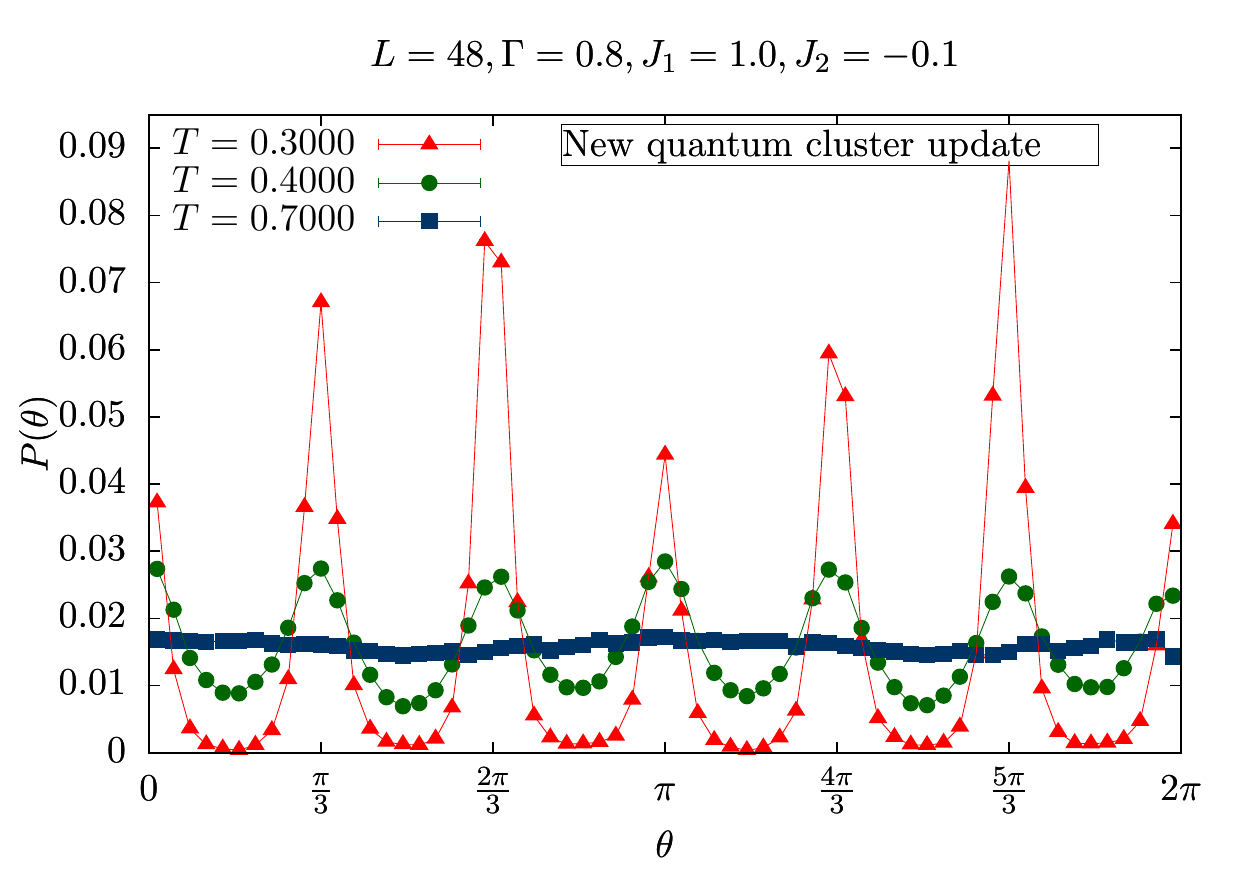}
  \caption{\label{48PF}Histograms of the phase of the estimator $\theta$ of the
complex three-sublattice order parameter $\psi$, obtained during
simulations performed using the new quantum cluster algorithm with $J_2=-0.1$,
$J_1=1$ and $\Gamma=0.8$. Note the
six clearly demarcated peaks at $2m\pi/6$ ($m=1.2 \dots 6$) in the low temperature state, characteristic of ferrimagnetic
three-sublattice order.}
\end{figure}

\begin{figure}[t]
  \includegraphics[width=7.6cm]{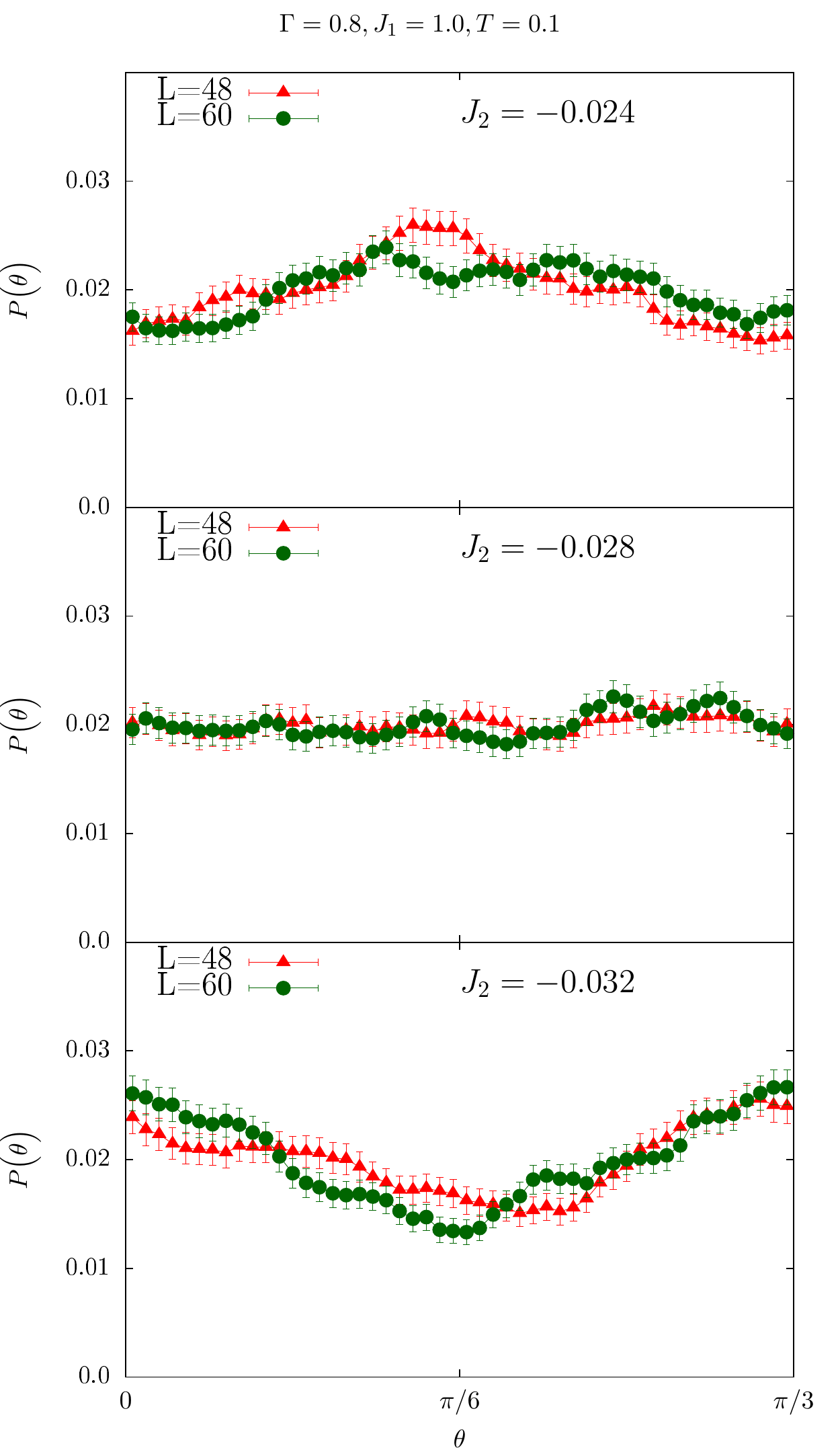}
  \caption{\label{1stOrd} Histograms of $\theta$, the phase of the estimator of
the complex three-sublattice order parameter $\psi$, obtained using
the new quantum cluster algorithm, allow us
to locate the transition between antiferromagnetic and ferrimagnetic three-sublattice ordered states of the triangular lattice transverse
field Ising antiferromagnet with $J_1=1$, $T=0.1$, and $\Gamma=0.8$. }
\end{figure}

If a ferromagnetic $J_2$ is present, one expects~\cite{DP_Landau} the nature of
the three-sublattice ordered state to change from antiferromagnetic three-sublattice
order to ferrimagnetic three-sublattice order beyond some critical value of $J_2$
beyond which the energetic effects of this coupling dominate over the quantum
fluctuations induced by $\Gamma$. This critical value will depend on the value of $\Gamma$ and $T$ (with $J_1$ fixed at $J_1=1$).
In Fig.~\ref{ordered} we see that the system has three sublattice order 
when $J_{2}=-0.1$, $J_{1}=1.0$ and $T=0.2$. This order is of the ferrimagnetic
kind. This is clear from
the histogram of $\theta$ obtained using the new 
quantum cluster algorithm. This histogram, shown in Fig.~\ref{48PF}, exhibits six well defined peaks at values $2m\pi/6$, corresponding  to ferrimagnetic three sublattice order. As in the antiferromagnetic phase, these  peaks become less pronounced on heating and are finally replaced by a flat histogram, which
presumably signals the onset of the 
the power-law ordered phase studied in Ref.~\onlinecite{DP_Landau} (we have not
studied this transition in any detail here).
As a further demonstration of the capabilities of the new quantum cluster algorithm,
we use it to track changes in the histogram of order parameter phase $\theta$ as a function
of $J_2$ at a fixed low temperature $T=0.1$, and pinpoint the location of this transition from antiferromagnetic three-sublattice order to ferrimagnetic three-sublattice order. As is clear
from the results shown in Fig.~\ref{1stOrd}, we obtain in this manner an accurate estimate
of $J_{2c} \approx -0.028(4)$ for the location of this transition
when $\Gamma=0.8$, $T=0.1$, and $J_1=1$.
\section{Discussion and Outlook}
\label{Outlook}

In the conventional quantum cluster algorithm, the cluster decomposition of a given
operator string is completely deterministic, although each cluster is randomly flipped
with probability $1/2$. In the small $\Gamma$ limit, the operator string
is very nearly completely diagonal, and the set of lattice sites belonging to a given
space-time cluster defines clusters that correspond precisely~\cite{Sandvik} to clusters
grown by the {\em classical} Swendsen-Wang~\cite{SW} cluster algorithm. Indeed, in
this sense, the conventional quantum cluster algorithm can
be viewed as a quantum generalization of the Swendsen-Wang cluster algorithm.

Thought of in this way, it is not surprising that it fails when the Ising exchange interactions
are frustrated, since the Swendsen-Wang algorithm is known to perform poorly
at low temperature in classical Ising models with frustrated interactions.
Since our new quantum cluster algorithm solves this problem, a natural question
that might be worth exploring in more detail is the classical limit of our
new quantum cluster algorithm. Preliminary analysis suggests that this classical limit 
is closely related to the classical cluster algorithm proposed by Zhang and Yang~\cite{Zhang} for the 
specific case of triangular lattice Ising model with nearest neighbour antiferromagnetic interactions, as well as the general plaquette algorithms explored by Coddington and Han~\cite{Coddington_Han}. It would 
be useful to explore this further in future work.

In Sec.~\ref{Method} and Sec.~\ref{Results}, we have already studied extensions of our quantum cluster algorithm to systems in which some of the exchange couplings are frustrated,
while others are unfrustrated, for instance, transverse-field Ising antiferromagnets
on the triangular lattice with nearest-neighbour antiferromagnetic interactions
and next-nearest neighbour ferromagnetic interactions. Additionally, we have also
tested the extension to the case in which both nearest and next-nearest neighbour
interactions are antiferromagnetic.
Another, far more nontrivial, extension would involve generalizations of our quantum cluster
approach to the simulation of models such as the quantum dimer model~\cite{MoessnerQDM,SchlittlerQDM}, in which
the Hilbert space consists of states that obey a set of local constraints. Preliminary
analysis suggests that this may also be possible, and is therefore worth exploring.

\section{Acknowledgements}
Our computational work was made possible by the computational resources
of the Department of Theoretical Physics of the Tata Institute of Fundamental
Research, as well as by computational resources funded by 
DST (India) grant DST-SR/S2/RJN-25/2006. The authors are grateful for
the hospitality of ICTP (Trieste) and IIT Guwahati during the drafting of part of this manuscript.

\bibliography{mybib}

\end{document}